\documentclass[12pt]{article}

\usepackage[
    textheight=22.5truecm,
    textwidth=16.8truecm,
    top=2.95truecm,
    left=2.15truecm
]{geometry}

\usepackage{
    amsmath,
    amsthm,
    amssymb,
    bm,
    braket,
    mathrsfs
}

\usepackage{
    graphicx,
    tikz,
    caption
}

\usepackage{
    color,
    titlesec,
    authblk,
    enumitem,
    titling,
    ascmac
}

\usepackage{
    hyperref,
    cite
}

\hypersetup{
    colorlinks,
    bookmarksopen,
    bookmarksnumbered,
    citecolor=black,
    linkcolor=black,
    linktocpage,
    pdfstartview=FitH,
    urlcolor=black
}

\numberwithin{equation}{section}

\renewcommand{\thefigure}{\thesection.\arabic{figure}}
\numberwithin{figure}{section}

\renewcommand{\thetable}{\thesection.\arabic{table}}
\numberwithin{table}{section}

\makeatletter
\long\def\@makefntext#1{
    \parindent 1em\noindent
    \@hangfrom{\hbox to 1.8em{\hss$^{\@thefnmark}$}}#1
}
\makeatother

\def\fnum@figure{\textbf{\figurename\nobreakspace\thefigure}}
\def\fnum@table{\textbf{\tablename\nobreakspace\thetable}}

\long\def\@makecaption#1#2{
    \vskip\abovecaptionskip
    \sbox\@tempboxa{\small #1. #2}
    \ifdim \wd\@tempboxa >\hsize
        \small #1. #2\par
    \else
        \global \@minipagefalse
        \hb@xt@\hsize{\hfil\box\@tempboxa\hfil}
    \fi
    \vskip\belowcaptionskip
}

\setlist[enumerate]{
    nosep,
    leftmargin=16pt
}

\setlist[itemize]{
    nosep,
    leftmargin=16pt
}

\renewcommand{\d}{\mathrm{d}}

\allowdisplaybreaks

\pretitle{
    \begin{flushright}
        \normalsize
        RIKEN-iTHEMS-Report-26
    \end{flushright}

    \begin{center}
        \fontsize{20pt}{24pt}\selectfont
}

\posttitle{
    \par
    \end{center}
    \vskip 1em
}

\title{
    A Lindbladian for holographic Brownian motion
}

\author{Daichi Takeda}

\affil{
    \it
    iTHEMS, RIKEN, Wako, Saitama 351-0198, Japan
}

\date{}

\begin{document}

\maketitle


\begin{abstract}
We derive a Lindbladian description of holographic Brownian motion in the high-temperature regime.
Starting from the influence functional for a trailing string endpoint, we identify the corresponding quantum master equation and prove that it is completely positive and trace-preserving.
We determine the coefficients of the Lindbladian explicitly for two holographic backgrounds: the BTZ black hole and the AdS\(_5\) black brane, restricting in the latter case to the endpoint fluctuation along the \(x^1\)-direction.
We then analyze the time evolution of phase-space moments, energy relaxation, and steady states.
\end{abstract}


\newpage
\tableofcontents
\newpage


\section{Introduction}
\label{sec: introduction}

In the AdS/CFT correspondence \cite{Maldacena:1997re,Gubser:1998bc,Witten:1998qj}, holographic Brownian motion \cite{deBoer:2008gu,Atmaja:2010uu}
provides a concrete framework for studying the stochastic dynamics of a heavy probe quark coupled to a strongly interacting thermal plasma.
In the gravitational description, the probe quark is represented by an open-string endpoint on the AdS boundary, following the standard holographic description of Wilson lines and external heavy quarks \cite{Maldacena:1998im,Rey:1998ik,Rey:1998bq,Brandhuber:1998bs,Danielsson:1998wt}.
Thermal fluctuations of the string worldsheet then induce dissipation and noise in the endpoint motion.
The resulting endpoint motion is classically described by a Langevin equation, whose dissipative and fluctuating coefficients are fixed by retarded and symmetric Green functions obtained from bulk analyses and obey the fluctuation-dissipation relation
\cite{CasalderreySolana:2006rq,Gubser:2006qh,Casalderrey-Solana:2007ahi,Son:2009vu,Giecold:2009cg,Casalderrey-Solana:2009ifi}.\footnote{
Related studies of holographic Brownian or Langevin dynamics in various contexts include
\cite{Gursoy:2010aa,Fischler:2012ff,Atmaja:2012jg,Atmaja:2013cxa,Banerjee:2013rca,Chakrabortty:2013kra,Giataganas:2013zaa,Sadeghi:2014lha,Banerjee:2015fed,Finazzo:2016mhm,Chakrabarty:2019aeu,Bu:2022oty,Zhou:2024oeg,Zhou:2024khk,Chowdhury:2025vai,Rajagopal:2026urq}.
}

The generating functional that determines real-time correlation functions is formulated in the Schwinger--Keldysh (SK) formalism.
At quadratic order, the generating functional is expressed as the exponential of a bilocal effective action determined by the corresponding retarded and symmetric Green functions.
Early holographic prescriptions for real-time thermal correlators were developed in \cite{Son:2002sd,Herzog:2002pc}.
More systematic prescriptions for SK generating functionals were subsequently developed in \cite{Skenderis:2008dg,Skenderis:2008dh,vanRees:2009rw,Crossley:2018tua}.
In the context of holographic Brownian motion, \cite{Bu:2021clf} constructed the SK generating functional for string-endpoint fluctuations around trailing-string backgrounds \cite{Herzog:2006gh,Gubser:2006bz} by using the prescription of \cite{Crossley:2018tua}.\footnote{
The analysis in \cite{Bu:2021clf} includes cubic fluctuations in addition to the quadratic influence functional considered in the present work.
}
The same quadratic generating functional was derived from the prescription of \cite{vanRees:2009rw} in \cite{NakamuraTakeda}.

If the external sources in an SK generating functional are themselves the dynamical variables of another system, the same generating functional can be viewed as the Feynman--Vernon influence functional for that system \cite{Feynman:1963fq}.
Indeed, the influence functional is obtained by holding the system histories arbitrarily fixed and integrating out the environment degrees of freedom.
In the Caldeira--Leggett model \cite{Caldeira:1982iu}, for example, one fixes the particle trajectory and integrates out the oscillator bath.
When the environment is a holographic CFT, this process is replaced by the evaluation of the bulk saddle-point action in the holographic SK geometry \cite{Jana:2020vyx}.
For the string endpoint, the resulting influence functional is precisely the string-endpoint generating functional constructed in \cite{Bu:2021clf,NakamuraTakeda}.

The Feynman--Vernon influence functional can be used to derive a quantum master equation for the reduced density matrix of the system.
In the high-temperature Markovian limit, this amounts to performing a local derivative expansion of the influence functional.
In the Caldeira--Leggett model, the first-order derivative expansion gives the familiar Caldeira--Leggett master equation \cite{Caldeira:1982iu}, but this equation is known not to preserve positivity in general.
Di\'osi's analysis \cite{Diosi:1993} showed that complete positivity requires keeping the derivative expansion to second order, which supplies an additional diffusion term in the master equation.\footnote{
It is worth mentioning that an exact non-Markovian master equation is derived in \cite{Hu:1991di} for the Caldeira--Leggett model.
}

Following this lesson, we construct a Lindbladian for holographic Brownian motion directly from the string-endpoint influence functional.\footnote{
See \cite{Ishii:2025} for the Lindblad dynamics for a holographic CFT. In the present paper, a holographic CFT is an environment.
}
We perform the high-temperature derivative expansion to second order, as in the Di\'osi master equation, and show that the resulting local quantum master equation is completely positive and trace-preserving.
We consider two holographic setups: the BTZ black hole and the AdS\(_5\) black brane.
For the AdS\(_5\) black brane, we restrict to the endpoint fluctuation along the \(x^1\)-direction and treat the corresponding one-dimensional sector.

After deriving the Lindbladian, we analyze its quantum Brownian dynamics.
For the free particle, we solve the time evolution of the phase-space moments and their late-time behavior.
We then add a harmonic potential as a simple confining force and compute the first and second moments in the steady state.
From these results, we determine the asymptotic energy expectation value and show that it approaches the classical equipartition value when the particle mass is sufficiently heavy in addition to that the black-hole temperature is high.

The rest of this paper is organized as follows.
\begin{itemize}
    \item
    In Section~\ref{sec: influence functional}, we perform the derivative expansion of the endpoint influence functional and determine the coefficients needed for the Lindbladian.
    \item
    In Section~\ref{sec: Lindbladian}, we promote the endpoint source to a dynamical variable and derive the Lindblad master equation, including the proof of complete positivity and trace preservation.
    \item
    In Section~\ref{sec: dynamics}, we study the dynamics generated by the Lindbladian for a free particle and for a particle in a harmonic potential, and analyze the corresponding energy relaxation and equipartition regime.
    \item
    In Section~\ref{sec: discussion}, we summarize the results and discuss open issues.
	\item
	In Appendix~\ref{app: 5D detail}, the detailed computation of Section~\ref{subsec: BB data} is shown.
	\item
	In Section~\ref{app: Wigner}, the formal solution to the Lindblad equation is provided in Wigner representation.
    \item
    In Appendix~\ref{app: finite cutoff}, keeping the location of the cutoff surface finite, we rederive the master equation for the BTZ background. Using the exact Green functions, we trace the initial slip and final dress.
\end{itemize}


\section{Derivative expansion of the influence functional}
\label{sec: influence functional}

In this section we extract the data that will be used to construct the Lindbladian in Section~\ref{sec: Lindbladian}.
We start from the quadratic endpoint influence functional written in terms of the retarded and symmetric Green functions, and perform its high-temperature derivative expansion.
We then determine these coefficients explicitly in two holographic backgrounds, the BTZ black hole and the AdS\(_5\) black brane.


\subsection{Data needed for the Lindbladian}
\label{subsec: data for Lindbladian}

We begin with reviewing the setup of the trailing string solution, the quadratic expansion around it, and the influence functional.
The data needed for constructing the Lindbladian will be explained at the end of this subsection.

\paragraph{String setup.}

We use the black-brane form of the metric
\begin{align}
\d s^2
=
\frac{r^2}{L^2}
\left[
-f(r)\d t^2
+
\sum_{i=1}^{d-1}(\d x^i)^2
\right]
+
\frac{L^2}{r^2f(r)}\d r^2,
\label{eq: general BB metric}
\end{align}
with \(L\) denoting the AdS radius.
The BTZ black hole and the AdS\(_5\) black brane correspond to different choices of \(d\) and \(f(r)\).
We place the endpoint of the open string at a cutoff surface \(r=r_c\), which is taken close to the AdS boundary.
Throughout this paper we restrict the string embedding to the \(t\), \(r\), and \(x^1\) directions,
\begin{align}
X^2=\cdots=X^{d-1}=0 .
\label{eq: ansatz}
\end{align}

The string dynamics is governed by the Nambu--Goto action
\begin{align}
S_{\mathrm{NG}}
=
-\frac{1}{2\pi\alpha^\prime}
\int \d^2\sigma\,
\sqrt{-\det h_{ab}},
\qquad
h_{ab}
=
G_{\mu\nu}(X)
\partial_aX^\mu
\partial_bX^\nu .
\label{eq: NG action}
\end{align}
We also introduce a local counterterm on the cutoff worldline to subtract the mass divergence as \(r_c\to\infty\):\footnote{\label{foot: divergent mass}
One may instead keep the large but finite endpoint inertial mass, which grows linearly with \(r_c\).
The framework developed below also applies to that scheme, and we will comment on it later.
}
\begin{align}
S_{\mathrm{ct}}
=
M_{\mathrm{ct}}
\int_{r=r_c}
\d t\,
\sqrt{-h_{tt}},
\qquad
M_{\mathrm{ct}} := \frac{L}{2\pi\alpha^\prime},
\label{eq: cutoff worldline counterterm}
\end{align}
where the endpoint trajectory at the cutoff surface is
\begin{align}
X^\mu_{\mathrm{end}}(t)
=
\left(
t,\,
r_c,\,
x^1(t),\,
0,\ldots,0
\right),
\label{eq: cutoff endpoint worldline}
\end{align}
and the induced metric on this worldline is
\begin{align}
h_{tt}
=
G_{\mu\nu}(X_{\mathrm{end}})
\frac{\d X^\mu_{\mathrm{end}}}{\d t}
\frac{\d X^\nu_{\mathrm{end}}}{\d t}
=-\frac{r_c^2}{L^2}f(r_c)
+
\frac{r_c^2}{L^2}\dot x^1(t)^2 .
\label{eq: cutoff worldline metric}
\end{align}

We take the classical string profile to be a trailing string,
\begin{align}
X^0=t,
\qquad
X^1=vt-\xi(r),
\label{eq: trailing string}
\end{align}
with the ansatz \eqref{eq: ansatz}.
This is the standard holographic description of a heavy quark moving through a thermal plasma \cite{Herzog:2006gh,Gubser:2006bz}.
The static string is included as the special case \(v=0\).

\paragraph{Trailing string and worldsheet geometry.}

For the trailing string profile \eqref{eq: trailing string}, the induced worldsheet metric in the static gauge \(\sigma^a=(t,r)\) is
\begin{align}
h_{tt}
=
\frac{r^2}{L^2}\left[v^2-f(r)\right],
\qquad
h_{tr}
=
-\frac{r^2}{L^2}v\xi^\prime(r),
\qquad
h_{rr}
=
\frac{L^2}{r^2f(r)}
+
\frac{r^2}{L^2}(\xi^\prime(r))^2 .
\label{eq: induced metric trailing}
\end{align}
The Nambu--Goto equation for \(\xi(r)\) has a conserved momentum because \(\xi\) is a cyclic variable:
\begin{align}
  S_\mathrm{NG} = -\frac{1}{2\pi\alpha^\prime}\int\d^2\sigma\sqrt{1 - \frac{v^2}{f(r)}+\frac{f(r)r^4}{L^4}\xi'^2}.
\end{align}
For the trailing string solution \cite{Herzog:2006gh,Gubser:2006bz}, which stretches to the horizon and does not come back to the AdS boundary, \(\xi\) is determined from
\begin{align}
\xi^\prime(r)
=
-
\frac{L^2 r_*^2 v}{r^2 f(r)}
\sqrt{
\frac{f(r)-v^2}{
r^4 f(r)-r_*^4 v^2
}
},\qquad
\xi(\infty)=0,
\label{eq: trailing profile derivative}
\end{align}
where \(r_*\) is defined via
\begin{align}
f(r_*)=v^2.
\label{eq: worldsheet horizon}
\end{align}

The off-diagonal component of the induced metric can be removed by introducing a worldsheet time coordinate \(\tau\) by
\begin{align}
\d\tau
=
\d t
+
\frac{h_{tr}}{h_{tt}}\d r ,
\label{eq: diagonal worldsheet time}
\end{align}
which satisfies, from \eqref{eq: induced metric trailing} and \eqref{eq: trailing profile derivative},
\begin{align}
\tau(t,r\to \infty) = t.
\label{eq: asymptotic tau}
\end{align}
Note that \eqref{eq: diagonal worldsheet time} allows an additional constant for \eqref{eq: asymptotic tau}, which we here fix to \(0\).
In the \((\tau,r)\) coordinates, the worldsheet metric becomes
\begin{align}
\d s_{\mathrm{ws}}^2
=
h_{tt}(r)\d\tau^2
+
\left(
h_{rr}
-
\frac{h_{tr}^2}{h_{tt}}
\right)
\d r^2
=: \tilde h_{\tau\tau}\d \tau^2 + \tilde h_{rr}\d r^2
 .
\label{eq: diagonal worldsheet metric}
\end{align}
The worldsheet horizon is located at \(r=r_*\), since we find \(\tilde h_{\tau\tau}\) and \(\tilde h_{rr}^{-1}\) vanish there.
The corresponding worldsheet Hawking temperature is given as
\begin{align}
T_*
=
\frac{1}{4\pi L^2}
\sqrt{
r_*^2 f^\prime(r_*)
\left[
r_*^2 f^\prime(r_*)+4r_*v^2
\right]
}.
\label{eq: worldsheet temperature general}
\end{align}

\paragraph{Expansion around the trailing string.}

We now turn on a fluctuation along the \(x^1\)-direction,
\begin{align}
X^1(t,r)
=
vt-\xi(r)+\Phi(t,r).
\label{eq: fluctuation around trailing string}
\end{align}
After the change of variables from \(t\) to the diagonal worldsheet time \(\tau\), we regard \(\Phi\) as a function of \((\tau,r)\).
The endpoint fluctuation is defined by the cutoff value
\begin{align}
q(\tau):=\Phi(\tau,r_c).
\label{eq: endpoint fluctuation definition}
\end{align}
Expanding the Nambu--Goto action around the trailing string, we write
\begin{align}
S_{\mathrm{NG}}
=
S_{\mathrm{NG}}^{(0)}
+
S_{\mathrm{NG}}^{(1)}
+
S_{\mathrm{NG}}^{(2)}
+
O(\Phi^3).
\label{eq: NG expansion}
\end{align}
The zeroth- and first-order terms are\footnote{
The surface contribution from the worldsheet horizon vanishes for the solutions on the bulk Schwinger--Keldysh contour, and thus it is omitted in \eqref{eq: NG linear}. See \cite{NakamuraTakeda}.
}
\begin{align}
S_{\mathrm{NG}}^{(0)}
&=
-\frac{1}{2\pi\alpha^\prime}
\int \d\tau\,\d r\,
\frac{r^2\sqrt{f(r)-v^2}}
{\sqrt{r^4f(r)-r_*^4v^2}},
\label{eq: NG zeroth}
\\
S_{\mathrm{NG}}^{(1)}
&=
F_\mathrm{fric}
\int \d\tau\,
q(\tau),
\label{eq: NG linear}
\end{align}
where \(F_{\mathrm{fric}}\) is defined as 
\begin{align}
F_{\mathrm{fric}}
&:=
-
\frac{1}{2\pi\alpha^\prime}
\frac{r_*^2v}{L^2}.
\label{eq: friction force definition}
\end{align}
This is nothing but the friction force that the probe quark feels from the bath CFT, since \eqref{eq: NG linear} is the linear response to the orbit \(x^1 = vt - \xi(r_c)\).

The quadratic term can be written in a covariant form on the diagonal worldsheet as
\begin{align}
S_{\mathrm{NG}}^{(2)}
=
-\frac{1}{4\pi\alpha^\prime}
\int \d\tau\,\d r\,
\mathcal K(r)
\sqrt{-\tilde h}\,
\tilde h^{ab}
\partial_a \Phi
\partial_b \Phi ,
\label{eq: quadratic NG action covariant}
\end{align}
where \(\tilde h_{ab}\) is given in \eqref{eq: diagonal worldsheet metric}, and we have defined
\begin{align}
\mathcal K(r)
=
\frac{
r^4 f(r)-r_*^4v^2
}{
L^2 r^2\left[f(r)-v^2\right]
}.
\label{eq: longitudinal fluctuation coefficient}
\end{align}
We can write \eqref{eq: quadratic NG action covariant} explicitly as
\begin{align}
S_{\mathrm{NG}}^{(2)}
&=
\frac{1}{4\pi\alpha^\prime}
\int \d\tau\,\d r
\left[
\mathcal Q(r)(\partial_\tau \Phi)^2
-
\mathcal P(r)(\partial_r \Phi)^2
\right],
\label{eq: quadratic NG action PQ}
\\
\mathcal Q(r)
&=
-\mathcal K(r)\sqrt{-\tilde h}\,\tilde h^{\tau\tau},
\qquad
\mathcal P(r)
=
\mathcal K(r)\sqrt{-\tilde h}\,\tilde h^{rr}.
\label{eq: PQ definition}
\end{align}

The counterterm \eqref{eq: cutoff worldline counterterm} is expanded in the same way.
At the cutoff surface, the endpoint trajectory is, from \eqref{eq: endpoint fluctuation definition}, given as
\begin{align}
x^1(\tau)
=
v\tau-\xi(r_c)+q(\tau),
\label{eq: cutoff endpoint fluctuation}
\end{align}
and hence we have
\begin{align}
\sqrt{-h_{tt}}
=
\frac{r_c}{L}\sqrt{f_c-v^2}
-
\frac{r_cv}{L\sqrt{f_c-v^2}}\dot q
-
\frac{r_cf_c}{2L\left(f_c-v^2\right)^{3/2}}
\dot q^{\,2}
+
O(q^3)
\qquad
\left(f_c:=f(r_c) \right).
\label{eq: counterterm sqrt expansion}
\end{align}
Therefore, the counterterm is expanded as
\begin{align}
S_{\mathrm{ct}}
&=
S_{\mathrm{ct}}^{(0)}
+
S_{\mathrm{ct}}^{(1)}
+
S_{\mathrm{ct}}^{(2)}
+
O(q^3),
\label{eq: ct expansion}
\\
S_{\mathrm{ct}}^{(0)}
&=
M_{\mathrm{ct}}
\frac{r_c}{L}\sqrt{f_c-v^2}
\int \d\tau,
\label{eq: ct zeroth}
\\
S_{\mathrm{ct}}^{(1)}
&=
-
M_{\mathrm{ct}}
\frac{r_cv}{L\sqrt{f_c-v^2}}
\int \d\tau\,\dot q,
\label{eq: ct linear}
\\
S_{\mathrm{ct}}^{(2)}
&=
-
\frac{M_{\mathrm{ct}}r_cf_c}
{2L\left(f_c-v^2\right)^{3/2}}
\int \d\tau\,
\dot q^{\,2}.
\label{eq: quadratic counterterm}
\end{align}
The linear counterterm is a total derivative and gives a local endpoint contribution.

\paragraph{Schwinger--Keldysh influence functional.}

We now construct the endpoint influence functional from the action on the Schwinger--Keldysh contour.
The two Lorentzian segments are denoted by \(f\) and \(b\).
The corresponding bulk fluctuations are \(\Phi_f\) and \(\Phi_b\), whose cutoff values define the endpoint fluctuations
\begin{align}
q_f(\tau):=\Phi_f(\tau,r_c),
\qquad
q_b(\tau):=\Phi_b(\tau,r_c).
\label{eq: SK endpoint fluctuations}
\end{align}
The real-time part of the influence functional is obtained from
\begin{align}
iS_{\mathrm{IF}}
=
iS_{\mathrm{NG}}[\Phi_f]
-
iS_{\mathrm{NG}}[\Phi_b]
+
iS_{\mathrm{ct}}[q_f]
-
iS_{\mathrm{ct}}[q_b].
\label{eq: IF from SK actions}
\end{align}
The zeroth-order contributions are independent of the endpoint fluctuation and cancel between \(f\) and \(b\).
The linear term from the Nambu--Goto action gives
\begin{align}
iS_{\mathrm{IF}}^{(1)}
&=
iF_{\mathrm{fric}}
\int \d\tau\,
\left[
q_f(\tau)-q_b(\tau)
\right]
=
iF_{\mathrm{fric}}
\int \d\tau\,
q_a(\tau),
\label{eq: linear IF drag}
\end{align}
The linear term from the counterterm is a total derivative and does not contribute to the master equation.

We next focus on the quadratic part.
We write the Fourier transform of the bulk fields as
\begin{align}
\Phi_s(\tau,r)
=
\int
\frac{\d\omega}{2\pi}\,
e^{-i\omega\tau}
\tilde \Phi_s(\omega,r),
\qquad
s=f,b,
\label{eq: SK Fourier mode}
\end{align}
with boundary values
\begin{align}
\tilde\Phi_s(\omega,r_c)=\tilde q_s(\omega).
\label{eq: SK boundary value}
\end{align}
On shell, the Nambu--Goto part of the quadratic action becomes
\begin{align}
S_{\mathrm{NG}}^{(2)}[\Phi_s]
=
-
\frac{1}{4\pi\alpha^\prime}
\int
\frac{\d\omega}{2\pi}\,
\tilde \Phi_s(-\omega,r)
\mathcal P(r)
\partial_r\tilde\Phi_s(\omega,r)
\bigg|_{r=r_c}
\qquad
\left(\mbox{for \(s = f,b\)} \right),
\label{eq: NG onshell boundary term}
\end{align}
together with the local subtraction from \(S_{\mathrm{ct}}^{(2)}\).

We introduce the \(r/a\) variables
\begin{align}
q_r
=
\frac{q_f+q_b}{2},
\qquad
q_a
=
q_f-q_b.
\label{eq: ra variables endpoint}
\end{align}
In this basis, the influence functional up to quadratic order takes the following form \cite{Bu:2021clf,NakamuraTakeda}:
\begin{align}
iS_{\mathrm{IF}}
=
iF_{\mathrm{fric}}
\int \d\tau\,
q_a(\tau)
-i
\int
\frac{\d\omega}{2\pi}\,
\tilde q_a(-\omega)
\tilde G_{\mathrm R}(\omega)
\tilde q_r(\omega)
-\frac{1}{2}
\int
\frac{\d\omega}{2\pi}\,
\tilde q_a(-\omega)
\tilde G_{\mathrm S}(\omega)
\tilde q_a(\omega).
\label{eq: IF GR GS with linear term}
\end{align}
Here \(\tilde G_{\mathrm R}(\omega)\) is the retarded Green function and \(\tilde G_{\mathrm S}(\omega)\) is the symmetric one.

The retarded kernel is extracted from the ingoing solution of the equation of motion,
\begin{align}
\partial_r
\left[
\mathcal P(r)\partial_r\varphi_\omega(r)
\right]
+
\omega^2\mathcal Q(r)\varphi_\omega(r)
=
0.
\label{eq: fluctuation EOM general}
\end{align}
We denote by \(\varphi_\omega^{\mathrm{in}}(r)\) the solution satisfying the ingoing boundary condition at the worldsheet horizon and normalized at infinity as
\begin{align}
\varphi_\omega^{\mathrm{in}}(\infty)=1.
\label{eq: ingoing mode normalization}
\end{align}
Near the AdS boundary, the ingoing solution has the expansion
\begin{align}
\varphi_\omega^{\mathrm{in}}(r)
=
1
+
\frac{L^4\omega^2}{2(1-v^2)r^2}
+
\frac{\varphi_\omega^{(3)}}{r^3}
+
O(r^{-4}),
\label{eq: ingoing near boundary expansion}
\end{align}
and the retarded kernel is written in terms of the coefficient \(\varphi_\omega^{(3)}\) as
\begin{align}
\tilde G_{\mathrm R}(\omega)
=
-\frac{3}{2\pi\alpha^\prime L^4\sqrt{1-v^2}}\,
\varphi_\omega^{(3)}.
\label{eq: GR from normalizable coefficient}
\end{align}
The term of \(O(r^{-2})\) gives divergent mass term, and the counterterm cancels it out at \(r_c\to \infty\).\footnote{
Here we formally use \(\int \d \tau \dot q(\tau)^2 = \int\d\omega \omega^2 \tilde q(-\omega)\tilde q(\omega)/(2\pi)\).
}
The symmetric kernel is fixed by the Schwinger--Keldysh construction and obeys the fluctuation-dissipation relation as
\begin{align}
\tilde G_{\mathrm S}(\omega)
=
-\operatorname{Im}\tilde G_{\mathrm R}(\omega)
\coth\frac{\omega}{2T_*},
\label{eq: FDR general}
\end{align}
where \(T_*\) is the worldsheet temperature defined in \eqref{eq: worldsheet temperature general}.

\paragraph{Data needed for the Lindbladian.}

The local Markovian limit relevant for the Lindblad equation is obtained by expanding the endpoint Green functions at frequencies small compared with the worldsheet temperature,
\begin{align}
\frac{\omega}{T_*}\ll1.
\label{eq: low frequency condition}
\end{align}
We parametrize the retarded kernel as
\begin{align}
\tilde G_{\mathrm R}(\omega)
=
-i\gamma\omega
+
\delta M\,\omega^2
+
O(\omega^3).
\label{eq: GR low frequency expansion}
\end{align}
The coefficient \(\gamma\) is the friction coefficient, meaning that it satisfies
\begin{align}
\gamma
=
-\frac{\partial F_{\mathrm{fric}}}{\partial v}
=
\frac{1}{2\pi\alpha^\prime L^2}
\left[
r_*^2
+
\frac{4r_*v^2}{f^\prime(r_*)}
\right],
\label{eq: gamma from friction force}
\end{align}
where we have used \eqref{eq: friction force definition} and \eqref{eq: worldsheet horizon} for the second equality.
This relation holds because \(-i\gamma\omega\) in \(\tilde G_\mathrm{R}\) contributes to the real-time classical equation of motion as \(\ddot q = - \gamma \dot q +\cdots \).
In Section~\ref{subsec: BTZ data} and \ref{subsec: BB data}, we explicitly confirm \eqref{eq: gamma from friction force}.

For the symmetric kernel, we write
\begin{align}
\tilde G_{\mathrm S}(\omega)
=
2D_{pp}
+
2\Delta_{qq}\omega^2
+
O(\omega^4).
\label{eq: GS low frequency expansion}
\end{align}
The absence of an \(O(\omega)\) term in \(G_{\mathrm S}\) follows from the symmetry of the noise kernel,
\begin{align}
\tilde G_{\mathrm S}(-\omega)=\tilde G_{\mathrm S}(\omega).
\label{eq: GS even}
\end{align}
This is consistent with the fluctuation-dissipation relation \eqref{eq: FDR general}, since both \(\operatorname{Im}G_{\mathrm R}(\omega)\) and \(\coth(\omega/2T_*)\) are odd functions of \(\omega\).

The coefficients \(\gamma\), \(\delta M\), \(D_{pp}\), and \(\Delta_{qq}\) are all real.
Their precise role in the Lindblad generator will be fixed in Section~\ref{sec: Lindbladian}.
The remainder of this section evaluates the coefficients in \eqref{eq: GR low frequency expansion} and \eqref{eq: GS low frequency expansion} for the BTZ black hole and the AdS\(_5\) black brane.


\subsection{BTZ}
\label{subsec: BTZ data}

\paragraph{Background data.}

We first specialize the general setup to the BTZ black hole.
In the notation of \eqref{eq: general BB metric}, this corresponds to
\begin{align}
d=2,
\qquad
f(r)
=
1-\frac{r_+^2}{r^2},
\qquad
T
=
\frac{r_+}{2\pi L^2}.
\label{eq: BTZ background data}
\end{align}
Using the general condition \eqref{eq: worldsheet horizon}, the worldsheet-horizon radius is calculated as
\begin{align}
r_*
=
\frac{r_+}{\sqrt{1-v^2}}.
\label{eq: BTZ worldsheet horizon}
\end{align}
The worldsheet temperature follows from \eqref{eq: worldsheet temperature general} as
\begin{align}
T_*
=
\frac{r_*\sqrt{1-v^4}}{2\pi L^2}
=
T\sqrt{1+v^2}.
\label{eq: BTZ worldsheet temperature}
\end{align}

\paragraph{Exact Green functions.}

For compactness, we introduce the dimensionless radial coordinate
\begin{align}
\rho(r)
:=
\sqrt{
\frac{r^2+v^2r_*^2}
{r_*^2(1+v^2)}
},
\label{eq: BTZ rho definition}
\end{align}
so that the worldsheet horizon is located at
\begin{align}
\rho(r_*)=1.
\label{eq: BTZ rho horizon}
\end{align}
The ingoing solution of \eqref{eq: fluctuation EOM general} is given exactly by
\begin{align}
\varphi_\omega^{\mathrm{in}}(r)
=
\left(
1
-
\frac{i\nu}{\rho(r)}
\right)
\left(
\frac{\rho(r)-1}{\rho(r)+1}
\right)^{-i\nu/2},
\qquad
\nu
:=
\frac{\omega}{2\pi T_*}.
\label{eq: BTZ exact ingoing solution}
\end{align}
This solution satisfies the ingoing boundary condition at the worldsheet horizon and is normalized as
\begin{align}
\varphi_\omega^{\mathrm{in}}(r\to\infty)=1.
\label{eq: BTZ exact solution normalization}
\end{align}
Its near-boundary expansion is
\begin{align}
\varphi_\omega^{\mathrm{in}}(r)
&=
1
+
\frac{L^4\omega^2}{2(1-v^2)r^2}
+
\frac{\varphi_\omega^{(3)}}{r^3}
+
O(r^{-4}),
\label{eq: BTZ ingoing expansion}
\\
\varphi_\omega^{(3)}
&=
\frac{i}{3}
\left(
\nu+\nu^3
\right)
r_*^3(1+v^2)^{3/2}.
\label{eq: BTZ normalizable coefficient}
\end{align}
Using \eqref{eq: GR from normalizable coefficient}, the Green functions are given as
\begin{align}
\tilde G_{\mathrm R}^{\mathrm{BTZ}}(\omega)
&=
-\frac{i}{2\pi\alpha^\prime}
\left[
\frac{r_*^2(1+v^2)}{L^2(1-v^2)}\omega
+
\frac{L^2}{(1-v^2)^2}\omega^3
\right].
\label{eq: BTZ exact GR}
\\
\tilde G_{\mathrm S}^{\mathrm{BTZ}}(\omega)
&=
\frac{1}{2\pi\alpha^\prime}
\left[
\frac{r_*^2(1+v^2)}{L^2(1-v^2)}\omega
+
\frac{L^2}{(1-v^2)^2}\omega^3
\right]
\coth\frac{\omega}{2T_*}.
\label{eq: BTZ exact GS}
\end{align}

\paragraph{Data needed for the Lindbladian.}

Expanding \eqref{eq: BTZ exact GR} and \eqref{eq: BTZ exact GS} at small \(\omega/T_*\), and comparing with \eqref{eq: GR low frequency expansion} and \eqref{eq: GS low frequency expansion}, we obtain the BTZ data needed for the Lindbladian:
\begin{align}
\gamma_{\mathrm{BTZ}}
&=
\frac{2\pi (LT_*^\mathrm{BTZ})^2}{\alpha^\prime(1-v^2)^2},
\qquad
\delta M_{\mathrm{BTZ}}
=
0,
\label{eq: BTZ Lindblad data 1}
\\
D_{pp}^{\mathrm{BTZ}}
&=
\gamma_{\mathrm{BTZ}}T_*^\mathrm{BTZ},
\qquad
\Delta_{qq}^{\mathrm{BTZ}}
=
\frac{\gamma_{\mathrm{BTZ}}}{12T_*^\mathrm{BTZ}}\left(1 + \frac{3}{\pi^2} \right),
\label{eq: BTZ Lindblad data 2}
\end{align}
where \(T_*^\mathrm{BTZ}\) is the worldsheet temperature in the BTZ case, namely \eqref{eq: BTZ worldsheet temperature}.
We also see that the friction coefficient satisfies \eqref{eq: gamma from friction force}.


\subsection{AdS\(_5\) black brane}
\label{subsec: BB data}

\paragraph{Background data.}

We next specialize the general setup to the AdS\(_5\) black brane.
In the notation of \eqref{eq: general BB metric}, this corresponds to
\begin{align}
d=4,
\qquad
f(r)
=
1-\frac{r_h^4}{r^4},
\qquad
T
=
\frac{r_h}{\pi L^2}.
\label{eq: BB background data}
\end{align}
Using the general condition \eqref{eq: worldsheet horizon}, the worldsheet-horizon radius is calculated as
\begin{align}
r_*
=
\frac{r_h}{(1-v^2)^{1/4}}.
\label{eq: BB worldsheet horizon}
\end{align}
The worldsheet temperature follows from \eqref{eq: worldsheet temperature general} as
\begin{align}
T_*
=
\frac{r_*\sqrt{1-v^2}}{\pi L^2}
=
(1-v^2)^{1/4}T
.
\label{eq: BB worldsheet temperature}
\end{align}
In this paper, we consider only the longitudinal fluctuation along the \(x^1\)-direction.

\paragraph{Low-frequency ingoing solution.}

For the AdS\(_5\) black brane, the fluctuation equation \eqref{eq: fluctuation EOM general} becomes, up to an overall constant factor,
\begin{align}
\partial_r
\left[
\frac{r^4-r_*^4}{L^4}
\partial_r\varphi_\omega(r)
\right]
+
\frac{\omega^2}{1-v^2}
\frac{r^4}{r^4-r_*^4}
\varphi_\omega(r)
=
0.
\label{eq: BB radial equation}
\end{align}
We introduce
\begin{align}
x
:=
\frac{r}{r_*},
\qquad
\nu
:=
\frac{\omega L^2}{r_*\sqrt{1-v^2}}
=
\frac{\omega}{\pi T_*},
\label{eq: BB dimensionless variables}
\end{align}
so that \eqref{eq: BB radial equation} becomes
\begin{align}
\frac{\d}{\d x}
\left[
(x^4-1)
\frac{\d\varphi_\omega}{\d x}
\right]
+
\nu^2
\frac{x^4}{x^4-1}
\varphi_\omega
=
0.
\label{eq: BB dimensionless radial equation}
\end{align}
We construct the ingoing solution perturbatively as
\begin{align}
\varphi_\omega^{\mathrm{in}}(x)
=
\psi_0(x)
+
\nu\psi_1(x)
+
\nu^2\psi_2(x)
+
\nu^3\psi_3(x)
+
O(\nu^4),
\label{eq: BB psi expansion}
\end{align}
with the UV normalization
\begin{align}
\psi_0(\infty)=1,
\qquad
\psi_n(\infty)=0
\qquad
(n\geq1).
\label{eq: BB psi UV normalization}
\end{align}
The ingoing condition at the worldsheet horizon is imposed as
\begin{align}
\varphi_\omega^{\mathrm{in}}(x)
\sim
\mathcal N(\nu)(x-1)^{-i\nu/4},
\qquad
\mathcal N(\nu)
=
1+\nu n_1+\nu^2n_2+\nu^3n_3+O(\nu^4).
\label{eq: BB horizon normalization}
\end{align}
This horizon normalization fixes the integration constants order by order.
The details of this construction are given in Appendix~\ref{app: 5D detail}.
The resulting large-\(x\) expansion is
\begin{align}
\varphi_\omega^{\mathrm{in}}(x)
&=
1
+
\frac{\nu^2}{2x^2}
+
\left[
\frac{i}{3}\nu
-
\frac{1}{3}\nu^2
+
ic_3\nu^3
+
O(\nu^4)
\right]
x^{-3}
+
O(x^{-4}),
\label{eq: BB ingoing large x}
\\
c_3
&:=
\frac{\pi}{12}
-
\frac{1}{6}\log2.
\label{eq: BB c3}
\end{align}
Then, the normalizable coefficient in \eqref{eq: ingoing near boundary expansion} reads
\begin{align}
\varphi_\omega^{(3)}
=
\frac{L^6}{(1-v^2)^{3/2}}
\left[
\frac{i\pi^2T_*^2}{3}\omega
-
\frac{\pi T_*}{3}\omega^2
+
ic_3\omega^3
+
O(\omega^4)
\right].
\label{eq: BB normalizable coefficient}
\end{align}

\paragraph{Data needed for the Lindbladian.}

Using \eqref{eq: GR from normalizable coefficient}, the retarded kernel becomes
\begin{align}
\tilde G_{\mathrm R}^{\mathrm{5D}}(\omega)
=
-i\gamma_{\mathrm{5D}}\omega
+
\delta M_{\mathrm{5D}}\omega^2
-
i\kappa_{\mathrm{5D}}\omega^3
+
O(\omega^4),
\label{eq: 5D GR expansion}
\end{align}
where comparison with \eqref{eq: GR low frequency expansion} gives
\begin{align}
\gamma_{\mathrm{5D}}
=
\frac{\pi L^2(T_*^{\mathrm{5D}})^2}{2\alpha^\prime(1-v^2)^2},
\qquad
\delta M_{\mathrm{5D}}
=
\frac{\gamma_{\mathrm{5D}}}{\pi T_*^{\mathrm{5D}}},
\qquad
\kappa_{\mathrm{5D}}
=
\frac{3c_3}{\pi^2(T_*^{\mathrm{5D}})^2}
\gamma_{\mathrm{5D}}.
\label{eq: 5D GR coefficients}
\end{align}
Here \(T_*^{\mathrm{5D}}\) denotes the worldsheet temperature given in \eqref{eq: BB worldsheet temperature}.
In addition, the friction coefficient satisfies \eqref{eq: gamma from friction force}.

The symmetric kernel follows from \eqref{eq: FDR general} as
\begin{align}
\tilde G_{\mathrm S}^{\mathrm{5D}}(\omega)
=
\left[
\gamma_{\mathrm{5D}}\omega
+
\kappa_{\mathrm{5D}}\omega^3
+
O(\omega^5)
\right]
\coth\frac{\omega}{2T_*^{\mathrm{5D}}}.
\label{eq: 5D GS expansion before Markov}
\end{align}
Expanding this expression at small \(\omega/T_*^{\mathrm{5D}}\) and comparing with \eqref{eq: GS low frequency expansion}, we obtain
\begin{align}
D_{pp}^{\mathrm{5D}}
=
\gamma_{\mathrm{5D}}T_*^{\mathrm{5D}},
\qquad
\Delta_{qq}^{\mathrm{5D}}
=
\frac{\gamma_{\mathrm{5D}}}{12T_*^{\mathrm{5D}}}
\left(
1
+
\frac{3}{\pi}
-
\frac{6\log2}{\pi^2}
\right).
\label{eq: 5D noise coefficients}
\end{align}


\section{A Lindbladian for holographic Brownian motion}
\label{sec: Lindbladian}

\begin{table}[t]
\centering
\begin{tabular}{c|c|c|c|c|c}
&
\(T_*\)
&
\(\gamma\)
&
\(\delta M\)
&
\(D_{pp}\)
&
\(\Delta_{qq}\)
\\
\hline
&&&&&
\\[-8pt]
BTZ
&
\(\displaystyle \sqrt{1+v^2}\,T\)
&
\(\displaystyle
\frac{2\pi L^2T_*^2}
{\alpha^\prime(1-v^2)^2}
\)
&
\(0\)
&
\(\displaystyle
\gamma T_*
\)
&
\(\displaystyle
\frac{\gamma}{12T_*}\left(1+\frac{3}{\pi^2} \right)
\)
\\[12pt]
5D
&
\(\displaystyle (1-v^2)^{1/4}T\)
&
\(\displaystyle
\frac{\pi L^2T_*^2}
{2\alpha^\prime(1-v^2)^2}
\)
&
\(\displaystyle
\frac{\gamma}{\pi T_*}
\)
&
\(\displaystyle
\gamma T_*
\)
&
\(\displaystyle
\frac{\gamma}{12T_*}\left(1+\frac{3}{\pi}
-\frac{6\log2}{\pi^2} \right)
\)
\end{tabular}
\caption{
Summary of the coefficients entering the Markovian influence functional.
}
\label{tab: Markov coefficients}
\end{table}

We now promote the endpoint fluctuation \(q\) to the dynamical coordinate of a quantum particle and derive the corresponding Markovian master equation.
The coefficients necessary for the master equation are summarized in Table~\ref{tab: Markov coefficients}.
Throughout this section we use the background-independent notation \eqref{eq: GR low frequency expansion} and \eqref{eq: GS low frequency expansion}.


\subsection{The Lindblad master equation}
\label{subsec: Lindblad}

We first make the endpoint fluctuation \(q\) dynamical by adding the action of a relativistic particle moving in the flat CFT metric,
\begin{align}
S_{\mathrm p}
=
-m
\int \d\tau\,
\sqrt{
1-\dot x^2
}.
\label{eq: relativistic particle action}
\end{align}
As we did for the bulk action, we expand this action around the uniform trajectory,
\begin{align}
x(\tau)
=
v\tau+q(\tau).
\label{eq: particle fluctuation}
\end{align}
Note that we have already taken \(r_c\to\infty\) to utilize \(\xi(\infty) = 0\) (see \eqref{eq: trailing profile derivative}).
Then we obtain
\begin{align}
S_{\mathrm p}
&=
S_{\mathrm p}^{(0)}
+
S_{\mathrm p}^{(1)}
+
S_{\mathrm p}^{(2)}
+
O(q^3),
\label{eq: particle action expansion}
\\
S_{\mathrm p}^{(0)}
&=
-m\sqrt{1-v^2}
\int \d\tau,
\label{eq: particle action zeroth}
\\
S_{\mathrm p}^{(1)}
&=
\frac{mv}{\sqrt{1-v^2}}
\int \d\tau\,\dot q,
\label{eq: particle action linear}
\\
S_{\mathrm p}^{(2)}
&=
\frac{M_q}{2}
\int \d\tau\,
\dot q^{\,2},
\qquad
M_q
:=
\frac{m}{(1-v^2)^{3/2}} .
\label{eq: particle action quadratic}
\end{align}
The zeroth-order contribution cancels between the \(f\) and \(b\) segments.
The first-order contribution is a surface term and will not affect the Lindblad generator.
Thus the system action relevant below is the nonrelativistic free-particle action\footnote{
This definition corresponds to using the renormalized inertial mass of the endpoint particle.
Alternatively, as mentioned in footnote~\ref{foot: divergent mass}, one may keep the large but finite endpoint inertial mass before subtracting the cutoff dependence.
This can be implemented by choosing \(m\) such that \(M_q\) is the unrenormalized inertial mass.
The derivation below is unchanged in that scheme.
See Appendix~\ref{app: finite cutoff} for the case of finite \(r_c\).
}
\begin{align}
S_{\mathrm p}^{(2)}
=
\frac{M_q}{2}
\int \d\tau\,
\dot q^{\,2}.
\label{eq: free particle action}
\end{align}

In addition, the uniform motion \(x=v\tau\) must be a classical trajectory for the dynamical endpoint \(x^1(\tau)\).
The string produces the friction force \(F_{\mathrm{fric}}\) defined in \eqref{eq: friction force definition}, and this force must be balanced by an external drag force
\begin{align}
F_{\mathrm{drag}}
:=
-F_{\mathrm{fric}}.
\label{eq: drag force definition}
\end{align}
We introduce this force through the relativistic worldline coupling with an external Maxwell field
\begin{align}
S_{\mathrm{ext}}
=
e
\int
A_\mu(x)\,\d x^\mu .
\label{eq: relativistic external coupling}
\end{align}
For a constant electric field in the \(x^1\)-direction, we choose
\begin{align}
A_t (x) = Ex^1,
\qquad
A_1= \cdots =  A_{d-1}=0,
\qquad
eE=F_{\mathrm{drag}}.
\label{eq: constant electric field gauge}
\end{align}
On the trajectory \(x^1=v\tau+q(\tau)\), this gives
\begin{align}
S_{\mathrm{ext}}
=
F_{\mathrm{drag}}
\int \d\tau\,
\left[
v\tau+q(\tau)
\right].
\label{eq: external force action full}
\end{align}
The first term is independent of \(q\) and cancels between the two Schwinger--Keldysh segments.
On the Schwinger--Keldysh contour, this contributes
\begin{align}
iS_{\mathrm{ext}}[q_f]
-
iS_{\mathrm{ext}}[q_b]
=
iF_{\mathrm{drag}}
\int \d\tau\,
q_a(\tau)
=
-iF_{\mathrm{fric}}
\int \d\tau\,
q_a(\tau),
\label{eq: external force SK contribution}
\end{align}
which cancels the linear friction term \eqref{eq: linear IF drag}.
Thus, after this cancellation, the Lindblad generator is determined by the quadratic influence functional.

Before taking the Markovian limit, it is useful to write the influence functional in real time.
We define the time-domain kernels by
\begin{align}
G_{\mathrm R}(\tau)
=
\int
\frac{\d\omega}{2\pi}\,
e^{-i\omega\tau}
\tilde G_{\mathrm R}(\omega),
\qquad
G_{\mathrm S}(\tau)
=
\int
\frac{\d\omega}{2\pi}\,
e^{-i\omega\tau}
\tilde G_{\mathrm S}(\omega).
\label{eq: time domain kernels}
\end{align}
By these, the influence functional \eqref{eq: IF GR GS with linear term}, after the cancellation of the linear term, is written as
\begin{align}
iS_{\mathrm{IF}}[q_f,q_b]
&:=
-i
\int \d\tau\,\d\tau^\prime\,
q_a(\tau)
G_{\mathrm R}(\tau-\tau^\prime)
q_r(\tau^\prime)
-\frac{1}{2}
\int \d\tau\,\d\tau^\prime\,
q_a(\tau)
G_{\mathrm S}(\tau-\tau^\prime)
q_a(\tau^\prime).
\label{eq: non-Markov IF time domain}
\end{align}
The reduced density matrix at time \(t\) is formally represented by the Schwinger--Keldysh path integral as
\begin{align}
\rho(Q_f,Q_b; t)
=
\int
\mathcal D q_f\,\mathcal D q_b\,
\exp
\left[
iS_{\mathrm p}^{(2)}[q_f]
-
iS_{\mathrm p}^{(2)}[q_b]
+
iS_{\mathrm{IF}}[q_f,q_b]
\right]
\rho_0(q_f(0),q_b(0)),
\label{eq: reduced density matrix path integral}
\end{align}
where \(Q_f\) and \(Q_b\) are the final coordinates of the wave function.

In the operator formalism, we may use the free Hamiltonian
\begin{align}
H_0
=
\frac{p^2}{2M_q}
\qquad
\left(\mbox{\(p\): the conjugate momentum of \(q\)} \right)
\label{eq: free Hamiltonian H0}
\end{align}
as the unperturbed Hamiltonian and regard the influence functional as the time-evolution operator in the interaction picture.
Introducing left and right actions on the density matrix,
\begin{align}
q_L\rho
:=
q\rho,
\qquad
q_R\rho
:=
\rho q,
\qquad
p_L\rho
:=
p\rho,
\qquad
p_R\rho
:=
\rho p,
\label{eq: left right q action}
\end{align}
we define the corresponding \(r/a\) combinations by
\begin{align}
q_a
=
q_L-q_R,
\qquad
q_r
=
\frac{q_L+q_R}{2},
\qquad
p_a
=
p_L-p_R,
\qquad
p_r
=
\frac{p_L+p_R}{2}
.
\label{eq: operator ra variables}
\end{align}
Similarly, we define\footnote{
For the free Hamiltonian \(H_0\), the momentum operators \(p_L(\tau)\) and \(p_R(\tau)\) are time independent.
The following derivation, however, does not need \(\dot p = 0\) and applies to the case with a potential like Section~\ref{subsec: harmonic oscillator}.
}
\begin{align}
p_L\rho
:=
p\rho,
\qquad
p_R\rho
:=
\rho p .
\label{eq: left right p action}
\end{align}
In this notation, \eqref{eq: reduced density matrix path integral} is written as
\begin{align}
\rho_I(t)
=
\mathcal T
\exp
\left[
iS_{\mathrm{IF}}
\left[
q_L,q_R
\right]
\right]
\rho(0),
\label{eq: non-Markov interaction picture evolution}
\end{align}
where \(q_L\) and \(q_R\) are now operators that evolve with \(H_0\), and \(\mathcal{T}\) is the time-ordering symbol.

We now take the local Markovian limit.
In doing so, we discard surface terms associated with integrations by parts, which do not contribute to the master equation (see Appendix~\ref{app: finite cutoff} for the surface terms).
Then, the low-frequency expansion of the kernels can be converted directly into a derivative expansion in time.
Using \eqref{eq: GR low frequency expansion} and \eqref{eq: GS low frequency expansion}, the influence functional is approximated as
\begin{align}
iS_{\mathrm{IF}}
&\simeq
-i\gamma
\int \d\tau\,
q_a\dot q_r
-i\delta M
\int \d\tau\,
\dot q_a\dot q_r
-
\int \d\tau\,
\left[
D_{pp}q_a^2
+
\Delta_{qq}\dot q_a^{\,2}
\right]
\nonumber
\\
&=
-i\frac{\gamma}{2M_q}
\int \d\tau\,
\left(q_L-q_R\right)
\left(p_L+p_R\right)
-i\frac{\delta M}{2M_q^2}
\int \d\tau\,
\left(p_L-p_R\right)
\left(p_L+p_R\right)
\nonumber
\\
&\quad
-
\int \d\tau\,
\left[
D_{pp}
\left(q_L-q_R\right)^2
+
D_{qq}
\left(p_L-p_R\right)^2
\right],
\label{eq: Markovian IF operator form}
\end{align}
where in the second equality we have used
\begin{align}
\dot q_L
=
\frac{p_L}{M_q},
\qquad
\dot q_R
=
\frac{p_R}{M_q},
\label{eq: free operator relation}
\end{align}
which holds in the interaction picture, and introduced
\begin{align}
D_{qq}
:=
\frac{\Delta_{qq}}{M_q^2}.
\label{eq: Dqq definition}
\end{align}

Thus, taking the time-derivative of \eqref{eq: non-Markov interaction picture evolution}, we obtain
\begin{align}
\frac{\d\rho_I}{\d t}
&=
-i\frac{\delta M}{2M_q^2}[ p_I^2,\rho_I]
-\frac{i\gamma}{2M_q}
\left[
q_I,
\left\{
p_I,\rho_I
\right\}
\right]
-D_{pp}
\left[
q_I,
\left[
q_I,\rho_I
\right]
\right]
-D_{qq}
\left[
p_I,
\left[
p_I,\rho_I
\right]
\right],
\label{eq: Lindblad master equation in I}
\end{align}
where \(I\) emphasizes that the operator is in the interaction picture.
Returning to the Schr\"odinger picture, we can rewrite it as
\begin{align}
\frac{\d\rho}{\d t}
&=
-i
\left[
H_{\mathrm{eff}},
\rho
\right]
-\frac{i\gamma}{2M_q}
\left[
q,
\left\{
p,\rho
\right\}
\right]
-D_{pp}
\left[
q,
\left[
q,\rho
\right]
\right]
-D_{qq}
\left[
p,
\left[
p,\rho
\right]
\right],
\label{eq: Lindblad master equation}
\\
H_{\mathrm{eff}}
&=
\frac{p^2}{2M}
,
\qquad
\frac{1}{M}
:=
\frac{1}{M_q}
+
\frac{\delta M}{M_q^2}.
\label{eq: effective Hamiltonian}
\end{align}
Eq.~\eqref{eq: Lindblad master equation} is the Markovian master equation obtained from the holographic influence functional.
While the trace preservation (TP) is already evident from \eqref{eq: Lindblad master equation}, the complete positivity (CP) is not immediate.
Below, we show that \eqref{eq: Lindblad master equation} is CPTP.


\subsection{Complete positivity and trace preservation}
\label{subsec: CPTP}

We now rewrite \eqref{eq: Lindblad master equation} in Lindblad form
\cite{Gorini:1976cm,Lindblad:1975ef}.
Let
\begin{align}
F_1= q,
\qquad
F_2=p,
\label{eq: Lindblad basis operators}
\end{align}
and introduce the Kossakowski matrix:
\begin{align}
\mathsf C
=
\begin{pmatrix}
2D_{pp} & -i\lambda
\\
i\lambda & 2D_{qq}
\end{pmatrix},
\qquad
\lambda
:=
\frac{\gamma}{2M_q}.
\label{eq: Kossakowski matrix}
\end{align}
Then, \eqref{eq: Lindblad master equation} can be rewritten as
\begin{align}
\frac{\d\rho}{\d t}
&=
-i
\left[
H_{\mathrm L},
\rho
\right]
+
\sum_{i,j=1}^{2}
\mathsf C_{ij}
\left(
F_i\rho F_j
-
\frac{1}{2}
\left\{
F_jF_i,\rho
\right\}
\right),
\label{eq: Lindblad form C matrix}
\\
H_{\mathrm L}
&=
H_{\mathrm{eff}}
+
\frac{\gamma}{4M_q}
\left\{
q,p
\right\}.
\label{eq: Lindblad Hamiltonian shifted}
\end{align}

Since \(\mathsf C\) is Hermitian, it can be diagonalized.
If \(\mathsf C\) is positive semidefinite, the diagonalized form is exactly of the Lindbladian with non-negative coefficients.
Thus complete positivity is equivalent to
\begin{align}
\mathsf C \geq 0 .
\label{eq: Kossakowski positivity}
\end{align}
Since \(D_{pp}\geq0\) and \(D_{qq}\geq0\), this condition reduces to
\begin{align}
\det \mathsf C
=
4D_{pp}D_{qq}
-
\frac{\gamma^2}{4M_q^2}
\geq0.
\label{eq: Kossakowski determinant condition}
\end{align}
Using \(D_{pp}=\gamma T_*\) and \eqref{eq: Dqq definition}, and writing
\begin{align}
\Delta_{qq}
=
\frac{\gamma}{12T_*}B,
\label{eq: Deltaqq B definition CPTP}
\end{align}
we obtain
\begin{align}
\det \mathsf C
=
\frac{\gamma^2}{3M_q^2}
\left(
B
-
\frac{3}{4}
\right).
\label{eq: Kossakowski determinant B}
\end{align}
The values of \(B\) read off from Table~\ref{tab: Markov coefficients} are
\begin{align}
B_{\mathrm{BTZ}}
=
1+\frac{3}{\pi^2},
\qquad
B_{\mathrm{5D}}
=
1+\frac{3}{\pi}
-\frac{6\log2}{\pi^2},
\label{eq: B values CPTP}
\end{align}
and both make \eqref{eq: Kossakowski determinant B} positive.
Therefore, \eqref{eq: Lindblad master equation} is a Lindblad equation and generates CPTP time evolution.

The positivity of \(\mathsf C\) is a nontrivial consequence of the holographic calculation; once the background is fixed,  the dissipative and diffusive coefficients are tied together by the worldsheet dynamics rather than chosen independently.


\section{Dynamics of the Lindbladian}
\label{sec: dynamics}

We now study the dynamics generated by the Lindblad equation \eqref{eq: Lindblad master equation}.
We first analyze the free Brownian particle and show that the momentum sector relaxes to a thermal distribution at the worldsheet temperature, while the position continues to diffuse.
We then add a harmonic potential and determine the stationary covariance matrix.
These results will be used to discuss the approach to the classical equipartition value.
See Appendix~\ref{app: Wigner} for the formal solution to the Lindblad equation \eqref{eq: Lindblad master equation}.


\subsection{Free Brownian motion}
\label{subsec: free Brownian}

\paragraph{Stationary distribution.}

We first analyze the free Brownian dynamics generated by \eqref{eq: Lindblad master equation}.
Let us look for a stationary density matrix depending only on \(p\),
\begin{align}
\rho
=
\rho_{\mathrm{ss}}(p).
\label{eq: free stationary ansatz}
\end{align}
For this ansatz, \eqref{eq: Lindblad master equation} reduces to
\begin{align}
\frac{\gamma}{M_q}
\frac{\d}{\d p}
\left[
p\rho_{\mathrm{ss}}(p)
\right]
+
D_{pp}
\rho_{\mathrm{ss}}^{\prime\prime}(p)
=
0.
\label{eq: free stationary equation rho}
\end{align}
The normalizable solution in momentum space is
\begin{align}
\rho_{\mathrm{ss}}(p)
\propto
\exp
\left[
-\frac{\gamma}{2M_qD_{pp}}p^2
\right]
=
\exp
\left[
-\frac{p^2}{2M_qT_*}
\right]
= 
e^{-H_0/T_*},
\label{eq: free stationary momentum solution}
\end{align}
where we used \(D_{pp}=\gamma T_*\) in Table~\ref{tab: Markov coefficients}, and also \eqref{eq: free Hamiltonian H0}.
Thus the momentum distribution is thermal at the worldsheet temperature \(T_*\).
This stationary state is not normalizable in the full Hilbert space unless the spatial direction is compactified.

\paragraph{First moments.}

The first moments obey, from \eqref{eq: Lindblad master equation},
\begin{align}
\frac{\d}{\d\tau}
\langle q\rangle
=
\frac{1}{M}
\langle p\rangle,
\qquad
\frac{\d}{\d\tau}
\langle p\rangle
=
-2\lambda
\langle p\rangle,
\label{eq: free first moment equations}
\end{align}
with \(\lambda\) defined in \eqref{eq: Kossakowski matrix}.
Those are easily solved to be
\begin{align}
\langle p\rangle_\tau
=
e^{-2\lambda\tau}
\langle p\rangle_0,
\qquad
\langle q\rangle_\tau
=
\langle q\rangle_0
+
\frac{1-e^{-2\lambda\tau}}{2M\lambda}
\langle p\rangle_0.
\label{eq: free first moment solutions}
\end{align}
The momentum expectation value relaxes exponentially, while the position expectation value approaches a shifted constant.
The subscript \(0\) here and below indicates the initial value.

\paragraph{Covariance equations.}

We define the covariance matrix by
\begin{align}
X
:=
\begin{pmatrix}
q
\\
p
\end{pmatrix},
\qquad
\Sigma_{ij}
:=
\frac{1}{2}
\left\langle
X_iX_j+X_jX_i
\right\rangle
-
\left\langle X_i\right\rangle
\left\langle X_j\right\rangle .
\label{eq: covariance matrix definition}
\end{align}
We denote its components as
\begin{align}
Q(t)
:=
\Sigma_{qq}(t),
\qquad
C(t)
:=
\Sigma_{qp}(t),
\qquad
P(t)
:=
\Sigma_{pp}(t).
\label{eq: QCP definitions free}
\end{align}
From \eqref{eq: Lindblad master equation}, their time evolution is determined by
\begin{align}
\dot P
&=
-4\lambda P
+
2D_{pp},\qquad
\dot C
=
\frac{1}{M}P
-
2\lambda C,
\qquad
\dot Q
=
\frac{2}{M}C
+
2D_{qq}.
\label{eq: free covariance equation}
\end{align}
The first equation gives
\begin{align}
P(t)
=
P_\infty
+
\left(
P_0-P_\infty
\right)
e^{-4\lambda t},
\qquad
P_\infty
:=
\frac{D_{pp}}{2\lambda}
=
M_qT_*.
\label{eq: free P solution}
\end{align}
Substituting this into the equation for \(C\), we obtain
\begin{align}
C(t)
&=
C_\infty
+
\left(
C_0-C_\infty
\right)
e^{-2\lambda t}
+
\frac{P_0-P_\infty}{2M\lambda}
\left(
e^{-2\lambda t}
-
e^{-4\lambda t}
\right),
\label{eq: free C solution}
\\
C_\infty
&=
\frac{P_\infty}{2M\lambda}
=
\frac{M_q^2T_*}{M\gamma}.
\label{eq: free C infinity}
\end{align}
Finally, integrating the equation for \(Q\), we find
\begin{align}
Q(t)
&=
Q_0
+
2D_{\mathrm{pos}}t
+
\frac{C_0-C_\infty}{M\lambda}
\left(
1-e^{-2\lambda t}
\right)
+
\frac{P_0-P_\infty}{4M^2\lambda^2}
\left(
1-e^{-2\lambda t}
\right)^2,
\label{eq: free Q solution}
\\
D_{\mathrm{pos}}
&=
D_{qq}
+
\frac{C_\infty}{M}
=
D_{qq}
+
\frac{M_q^2T_*}{M^2\gamma}.
\label{eq: free position diffusion}
\end{align}
Thus the momentum sector relaxes to \(P_\infty=M_qT_*\), whereas the position sector continues to diffuse as
\begin{align}
Q(t)
=
2D_{\mathrm{pos}}t
+
O(1).
\label{eq: free Q late time}
\end{align}
This is why the free particle has no normalizable stationary state in the full Hilbert space.

\paragraph{Comparison with Langevin position diffusion.}
The position diffusion constant can be written as
\begin{align}
D_{\mathrm{pos}}
=
\frac{T_*}{\gamma}
\left[
\left(
\frac{M_q}{M}
\right)^2
+
\frac{\gamma D_{qq}}{T_*}
\right].
\label{eq: Dpos compared with Langevin}
\end{align}
Using \eqref{eq: effective Hamiltonian}, \eqref{eq: Dqq definition} and Table~\ref{tab: Markov coefficients}, we have
\begin{align}
\left(
\frac{M_q}{M}
\right)^2
=
1
+
O\left(
\frac{T_*}{M_q}
\right),
\qquad
\frac{\gamma D_{qq}}{T_*}
=
O\left(
\frac{T_*^2}{M_q^2}
\right).
\label{eq: Dpos correction estimate}
\end{align}
Therefore, in the heavy-particle regime \(T_*/M_q\ll1\), we have
\begin{align}
D_{\mathrm{pos}}
=
\frac{T_*}{\gamma}
\left[
1
+
O\left(
\frac{T_*}{M_q}
\right)
+
O\left(
\frac{T_*^2}{M_q^2}
\right)
\right].
\label{eq: Dpos Langevin regime}
\end{align}
Thus the standard Langevin diffusion constant \(T_*/\gamma\), as obtained in holographic Brownian motion \cite{deBoer:2008gu}, is recovered in the heavy-particle limit.


\subsection{Brownian motion with a harmonic potential}
\label{subsec: harmonic oscillator}

\paragraph{Setup.}

We next consider the static case \(v=0\) and add a harmonic potential for the endpoint.
Since \(F_{\mathrm{fric}}=0\) at \(v=0\), the constant external drag force introduced in \eqref{eq: drag force definition} is absent.
Instead, we use the same worldline coupling \eqref{eq: relativistic external coupling} to introduce a confining potential.
Choosing
\begin{align}
eA_t(x)
=
-\frac{k}{2}(x^1)^2,
\qquad
A_1=\cdots=A_{d-1}=0,
\label{eq: harmonic external gauge}
\end{align}
we obtain, for \(x^1=q(\tau)\),
\begin{align}
S_{\mathrm{ext}}
=
-\frac{k}{2}
\int \d\tau\,
q(\tau)^2 .
\label{eq: harmonic external action}
\end{align}
Thus the original and effective Hamiltonian become
\begin{align}
H_{0}
=
\frac{p^2}{2M_q}
+
\frac{k}{2}q^2
,\qquad
H_{\mathrm{eff}}
=
\frac{p^2}{2M}
+
\frac{k}{2}q^2 .
\label{eq: harmonic Hamiltonians}
\end{align}

\paragraph{First moments.}

The first moments obey
\begin{align}
\frac{\d}{\d\tau}
\langle q\rangle
=
\frac{1}{M}
\langle p\rangle,
\qquad
\frac{\d}{\d\tau}
\langle p\rangle
=
-k
\langle q\rangle
-
2\lambda
\langle p\rangle .
\label{eq: harmonic first moment equations}
\end{align}
Combining them, we obtain
\begin{align}
\frac{\d^2}{\d\tau^2}
\langle q\rangle
+
2\lambda
\frac{\d}{\d\tau}
\langle q\rangle
+
\frac{k}{M}
\langle q\rangle
=
0.
\label{eq: damped oscillator first moment}
\end{align}
Since \(\lambda>0\), the first moments always relax to
\begin{align}
\langle q\rangle_{\infty}
=
0,
\qquad
\langle p\rangle_{\infty}
=
0.
\label{eq: harmonic first moment steady}
\end{align}

\paragraph{Stationary covariance matrix.}

We use the same covariance variables \(Q,C,P\) as in \eqref{eq: QCP definitions free}.
Their time evolution is determined by
\begin{align}
\dot Q
=
\frac{2}{M}C
+
2D_{qq},
\qquad
\dot C
=
\frac{1}{M}P
-
kQ
-
2\lambda C,
\qquad
\dot P
=
-2kC
-
4\lambda P
+
2D_{pp}.
\label{eq: harmonic covariance equations}
\end{align}
Those can be written as
\begin{align}
\dot\Sigma
=
A\Sigma+\Sigma A^{\mathrm T}+2\mathcal D,
\qquad
A=
\begin{pmatrix}
0 & 1/M\\
-k & -2\lambda
\end{pmatrix},
\qquad
\mathcal D=
\begin{pmatrix}
D_{qq} & 0\\
0 & D_{pp}
\end{pmatrix}.
\label{eq: harmonic covariance Lyapunov}
\end{align}
The formal solution is then given by
\begin{align}
\Sigma(t)
=
e^{At}\Sigma(0)e^{A^{\mathrm T}t}
+
2\int_0^t \d s\,
e^{A(t-s)}
\mathcal D
e^{A^{\mathrm T}(t-s)} .
\label{eq: harmonic covariance formal solution}
\end{align}
For \(k>0\) and \(\lambda>0\), the eigenvalues of \(A\) have negative real parts.
Therefore the first term decays, the integral converges, and thus \(\Sigma(\tau)\) approaches the unique stationary solution.

The steady-state values are obtained by setting the left-hand sides of \eqref{eq: harmonic covariance equations} to zero.
The first equation gives
\begin{align}
C_{\infty}
=
-MD_{qq}.
\label{eq: harmonic Css}
\end{align}
The third equation then gives
\begin{align}
P_{\infty}
=
\frac{D_{pp}+kMD_{qq}}{2\lambda}
=
M_qT_*
+
\frac{kMD_{qq}}{2\lambda},
\label{eq: harmonic Pss}
\end{align}
where we have used \(D_{pp}=\gamma T_*\) from Table~\ref{tab: Markov coefficients} and the definition of \(\lambda\) in \eqref{eq: Kossakowski matrix}.
While we keep using the notation \(T_*\), we understand \(T_* = T\) since we now consider \(v=0\).
Finally, the second equation gives
\begin{align}
Q_{\infty}
&=
\frac{P_{\infty}}{Mk}
+
\frac{2\lambda C_\infty}{k}
=
\frac{M_qT_*}{Mk}
+
\frac{D_{qq}}{2\lambda}
+
\frac{2\lambda M}{k}D_{qq}.
\label{eq: harmonic Qss}
\end{align}
Thus the harmonic potential converts the position diffusion of the free particle into a finite stationary width.

From \eqref{eq: harmonic first moment steady}, those cumulants are equal to the corresponding moments.


\subsection{Equipartition and energy corrections}
\label{subsec: energy}

\paragraph{Two notions of energy.}

There are two natural quadratic Hamiltonians in the present Markovian description.
The original Hamiltonian is
\begin{align}
H_0
=
\frac{p^2}{2M_q}
+
\frac{k}{2}q^2,
\label{eq: oscillator H0 energy}
\end{align}
whereas the Hamiltonian with a thermal mass correction is
\begin{align}
H_{\mathrm{eff}}
=
\frac{p^2}{2M}
+
\frac{k}{2}q^2.
\label{eq: oscillator Heff energy}
\end{align}
We therefore evaluate both energies.\footnote{
We do not use \(H_{\mathrm L}\) in \eqref{eq: Lindblad Hamiltonian shifted} as the mechanical energy.
It arose when organizing \eqref{eq: Lindblad master equation} to \eqref{eq: Lindblad form C matrix}, and is not of the kinetic-plus-potential form. We therefore do not regard it as the mechanical energy of the particle.
}
Below, the free Brownian motion corresponds to \(k = 0\), and we set \(v=0\) when \(k\neq 0\).

\paragraph{Free particle.}

For the free particle, the reference energy is
\begin{align}
\left\langle H_0\right\rangle_t
=
\frac{1}{2M_q}
\left[
P(t)
+
\langle p\rangle_t^2
\right].
\label{eq: free H0 expectation}
\end{align}
Using \eqref{eq: free first moment solutions} and \eqref{eq: free P solution}, we obtain
\begin{align}
\left\langle H_0\right\rangle_{\infty}
=
\frac{P_\infty}{2M_q}
=
\frac{T_*}{2}.
\label{eq: free H0 equipartition}
\end{align}
Thus \(H_0\) satisfies the one-dimensional equipartition law exactly.
This is consistent with \eqref{eq: free stationary momentum solution}.

On the other hand, we similarly obtain
\begin{align}
\left\langle H_{\mathrm{eff}}\right\rangle_{\infty}
=
\frac{M_q}{2M}T_*
=
\frac{T_*}{2}
\left[
1+\frac{\delta M}{M_q}
\right],
\label{eq: free Heff equipartition correction}
\end{align}
where we have used \eqref{eq: effective Hamiltonian}.
Thus the two definitions agree at leading order in the heavy-particle regime, \(T_*/M_q \ll 1\).

\paragraph{Harmonic oscillator.}

We now use the stationary covariances obtained in \eqref{eq: harmonic Pss} and \eqref{eq: harmonic Qss}.
Since the first moments vanish at late times, the reference Hamiltonian \(H_0\) gives
\begin{align}
\left\langle H_0\right\rangle_{\infty}
=
K_{0,\infty}
+
V_{\infty},
\qquad
K_{0,\infty}
:=
\frac{P_\infty}{2M_q},
\qquad
V_{\infty}
:=
\frac{kQ_\infty}{2}.
\label{eq: oscillator H0 energy decomposition}
\end{align}
Substituting \eqref{eq: harmonic Pss} and \eqref{eq: harmonic Qss}, we obtain
\begin{align}
K_{0,\infty}
&=
\frac{T_*}{2}
+
\frac{kMD_{qq}}{4\lambda M_q},
\label{eq: oscillator K0 infinity}
\\
V_{\infty}
&=
\frac{M_q}{M}\frac{T_*}{2}
+
\frac{kD_{qq}}{4\lambda}
+
\lambda M D_{qq}.
\label{eq: oscillator V infinity}
\end{align}
Therefore \(\left\langle H_0\right\rangle_{\infty}\) is evaluated as
\begin{align}
\left\langle H_0\right\rangle_{\infty}
&=
\frac{T_*}{2}
\left[
1+\frac{M_q}{M}
\right]
+
\frac{kD_{qq}}{4\lambda}
\left[
1+\frac{M}{M_q}
\right]
+
\lambda M D_{qq}.
\label{eq: oscillator H0 energy}
\end{align}

Similarly, the effective Hamiltonian is decomposed as
\begin{align}
\left\langle H_{\mathrm{eff}}\right\rangle_{\infty}
=
K_{\mathrm{eff},\infty}
+
V_{\infty},
\qquad
K_{\mathrm{eff},\infty}
:=
\frac{P_\infty}{2M}.
\label{eq: oscillator Heff energy decomposition}
\end{align}
For the kinetic part, we have
\begin{align}
K_{\mathrm{eff},\infty}
=
\frac{M_q}{M}\frac{T_*}{2}
+
\frac{kD_{qq}}{4\lambda},
\label{eq: oscillator Keff infinity}
\end{align}
and thus we obtain
\begin{align}
\left\langle H_{\mathrm{eff}}\right\rangle_{\infty}
=
\frac{M_q}{M}T_*
+
\frac{kD_{qq}}{2\lambda}
+
\lambda M D_{qq}.
\label{eq: oscillator Heff energy}
\end{align}

\paragraph{Equipartition regime.}

We now examine the kinetic and potential contributions separately and confirm the equipartition law.
From Table~\ref{tab: Markov coefficients} and \eqref{eq: Dqq definition}, we have
\begin{align}
\frac{\delta M}{M_q}
=
O\left(
\frac{T_*}{M_q}
\right),
\qquad
\frac{\gamma D_{qq}}{T_*}
=
O\left(
\frac{T_*^2}{M_q^2}
\right).
\label{eq: energy small parameters}
\end{align}
The Markovian expansion also requires the oscillator frequency scale to satisfy
\begin{align}
\omega_{\mathrm{sys}}^2
\sim
\frac{k}{M_q}
\ll
T_*^2 .
\label{eq: oscillator high temperature condition}
\end{align}
Then \eqref{eq: oscillator K0 infinity}, \eqref{eq: oscillator Keff infinity}, and \eqref{eq: oscillator V infinity} give
\begin{align}
K_{0,\infty}
&=
\frac{T_*}{2}
\left[
1
+
O\left(
\frac{\omega_{\mathrm{sys}}^2}{T_*^2}
\right)
\right],
\label{eq: K0 equipartition regime}
\\
K_{\mathrm{eff},\infty}
&=
\frac{T_*}{2}
\left[
1
+
O\left(
\frac{T_*}{M_q}
\right)
+
O\left(
\frac{\omega_{\mathrm{sys}}^2}{T_*^2}
\right)
\right].
\label{eq: Keff equipartition regime}
\\
V_{\infty}
&=
\frac{T_*}{2}
\left[
1
+
O\left(
\frac{T_*}{M_q}
\right)
+
O\left(
\frac{\omega_{\mathrm{sys}}^2}{T_*^2}
\right)
\right],
\label{eq: V equipartition regime}
\end{align}
Thus, each approaches to the equipartition energy \(T_*/2\) in the heavy-particle and high-temperature regime,
\begin{align}\label{eq: heavy and high T}
  \omega_{\mathrm{sys}} \ll T_* \ll M_q.
\end{align}
Combining these results, both \(H_0\) and \(H_{\mathrm{eff}}\) approach \(T_*\) under \eqref{eq: heavy and high T}:
\begin{align}
\left\langle H_0\right\rangle_{\infty}
\sim
\left\langle H_{\mathrm{eff}}\right\rangle_{\infty}
\sim
T_*.
\label{eq: oscillator equipartition total}
\end{align}


\section{Discussion}
\label{sec: discussion}

In this paper we constructed a Lindbladian for holographic Brownian motion from the influence functional of the string endpoint.
After promoting the endpoint fluctuation to a dynamical variable, we performed the low-frequency (or high-temperature) Markovian expansion of the retarded and symmetric Green functions.
For the BTZ black hole, the coefficients were extracted from the exact ingoing solution, while for the AdS\(_5\) black brane they were obtained by solving the equation of motion perturbatively in \(\omega\).
The resulting master equation is \eqref{eq: Lindblad master equation}, for which the term \(\Delta_{qq}\omega^2\) is necessary to be CPTP.

We also studied the dynamics generated by this Lindbladian.
For the free particle, the momentum sector relaxes to a thermal distribution at the worldsheet temperature, while the position variance continues to grow diffusively.
The corresponding position diffusion constant reduces, in the heavy-particle limit, to the standard Langevin value.
With a harmonic potential at \(v=0\), the confining potential turns the position diffusion into a finite stationary width.
Evaluating the late-time energies associated with \(H_0\) and \(H_{\mathrm{eff}}\), we found that both approach the expected equipartition values in the regime \eqref{eq: heavy and high T}.
Thus the Markovian Lindblad description is consistent with the known Brownian coefficients and with the thermal behavior expected in its regime of validity.

The most direct extension is to include transverse fluctuations.
In the present work we restricted the AdS\(_5\) black-brane analysis to the longitudinal fluctuation along the \(x^1\)-direction of motion.
A full description of the heavy quark should contain one longitudinal and two transverse endpoint coordinates.
It would then be possible to derive a \((3+1)\)-dimensional Lindblad equation for the holographic heavy quark.
This would provide a holographic counterpart of the Lindblad descriptions proposed for real heavy quarks, for example, \cite{Akamatsu:2014qsa}.

Another direction is to understand whether the present Lindblad equation can be derived directly from a bulk operator description.
In the original analysis of holographic Brownian motion \cite{deBoer:2008gu}, the quadratic expansion of the Nambu--Goto string is quantized in the black-hole background, and the endpoint two-point function is obtained by imposing the thermal state associated with the worldsheet horizon.
By reformulating the present construction in that language, we may find the connection between the time evolution of the string state and the present Lindblad dynamics.

This question is also related to the thermodynamic interpretation.
Since we now have the quantum CPTP map, it admits the usual open-system notions of entropy production and second-law inequalities.
It would thus be beneficial to understand whether the entropy production of the endpoint Lindblad dynamics has a direct worldsheet interpretation.
For example, one may ask whether it can be related to energy, entropy, or information flow through the worldsheet horizon.
Such an interpretation would give a more geometric meaning to the second law in this holographic open-system setting.

Finally, it is important to go beyond the local quadratic approximation used in this paper.
The Markovian master equation was obtained by expanding the nonlocal influence functional at low frequency, and the string action itself was truncated at quadratic order in the fluctuation \(q\).
These two approximations are likely related in the ultraviolet.
Indeed, the non-Markovian influence functional discussed in Appendix~\ref{app: finite cutoff} already show a singular behavior.
This may reflect the fact that the quadratic approximation removes nonlinear mechanisms that would suppress sufficiently high-frequency fluctuations.
It would be helpful to incorporate the higher-order of the Nambu--Goto action and see whether or not this divergence problem is resolved.

\subsection*{Acknowledgement}
I thank Masataka Matsumoto and Shin Nakamura for discussions. Especially, they shared with me the idea of adding a potential term to the string endpoint, based on which Section~\ref{subsec: harmonic oscillator} was written. Also, Shin Nakamura advised me to see the equipartition law, which appears in Section \ref{subsec: energy}.
My work is supported by RIKEN Special Postdoctoral Researchers Program.


\appendix


\section{Low-frequency expansion of the ingoing solution on AdS\(_5\) black brane}
\label{app: 5D detail}

In this appendix we derive \eqref{eq: BB ingoing large x}.
We start from the dimensionless radial equation \eqref{eq: BB dimensionless radial equation},
\begin{align}
\frac{\d}{\d x}
\left[
(x^4-1)
\frac{\d\varphi_\omega}{\d x}
\right]
+
\nu^2
\frac{x^4}{x^4-1}
\varphi_\omega
=
0,
\label{eq: appB EOM}
\end{align}
where \(x\) and \(\nu\) are defined in \eqref{eq: BB dimensionless variables}.
We look for the solution normalized at the AdS boundary and ingoing at the worldsheet horizon.
For this purpose, we expand the solution as
\begin{align}
\varphi_\omega^{\mathrm{in}}(x)
=
\psi_0(x)
+
\nu\psi_1(x)
+
\nu^2\psi_2(x)
+
\nu^3\psi_3(x)
+
O(\nu^4),
\label{eq: app 5D psi expansion}
\end{align}
and impose the UV normalization
\begin{align}
\psi_0(\infty)=1,
\qquad
\psi_n(\infty)=0
\qquad
(n\geq1).
\label{eq: app 5D UV normalization}
\end{align}
The ingoing condition near the worldsheet horizon takes the form \eqref{eq: BB horizon normalization}:
\begin{align}
\varphi_\omega^{\mathrm{in}}(x)
\overset{x\to 1}{\to }
\mathcal N(\nu)(x-1)^{-i\nu/4},
\qquad
\mathcal N(\nu)
=
1+\nu n_1+\nu^2n_2+\nu^3n_3+O(\nu^4).
\label{eq: app 5D horizon normalization}
\end{align}
The factor \(\mathcal N(\nu)\) is fixed order by order.

It is useful to introduce the differential operator
\begin{align}
\mathcal L_x
:=
\frac{\d}{\d x}
\left[
(x^4-1)
\frac{\d}{\d x}
\right].
\label{eq: app 5D Lx}
\end{align}
Substituting \eqref{eq: app 5D psi expansion} into \eqref{eq: appB EOM}, we obtain
\begin{align}
\mathcal L_x\psi_0=0,\qquad
\mathcal L_x\psi_1=0,\qquad
\mathcal L_x\psi_2=-\frac{x^4}{x^4-1}\psi_0,\qquad
\mathcal L_x\psi_3=-\frac{x^4}{x^4-1}\psi_1 .
\label{eq: app 5D psi equations}
\end{align}
Matching the near-horizon expansion of \eqref{eq: app 5D psi expansion} with \eqref{eq: app 5D horizon normalization} fixes the integration constants.

The zeroth-order solution is fixed by \eqref{eq: app 5D UV normalization} and \eqref{eq: app 5D horizon normalization} as
\begin{align}
\psi_0(x)=1.
\label{eq: app 5D psi0}
\end{align}

At the first order, we need the homogeneous solution that vanishes at infinity \eqref{eq: app 5D UV normalization} and has the logarithmic singularity required by the ingoing condition \eqref{eq: app 5D horizon normalization}.
The solution is
\begin{align}
\psi_1(x)
&=
-iI_1(x),
\label{eq: app 5D psi1}
\\
I_1(x)
&:=
\int_\infty^x
\frac{\d y}{y^4-1}
=\frac{1}{4}\log\frac{x-1}{x+1} - \frac{1}{2}\arctan x + \frac{\pi}{4}
.
\label{eq: app 5D I1}
\end{align}
Expanding \eqref{eq: app 5D psi1} at large \(x\), we find
\begin{align}
\psi_1(x)
=
\frac{i}{3x^3}
+
O(x^{-7}).
\label{eq: app 5D psi1 large x}
\end{align}
Near the horizon, the same integral behaves as
\begin{align}
I_1(x)
=
\frac{1}{4}\log(x-1)
+
c_h
+
O(x-1),
\qquad
c_h
:=
\frac{\pi}{8}
-
\frac{1}{4}\log2.
\label{eq: app 5D I1 horizon}
\end{align}
Comparing \(\psi_1(x)\) with the expansion of \eqref{eq: app 5D horizon normalization} at \(O(\nu)\) , we obtain
\begin{align}
n_1
=
-ic_h.
\label{eq: app 5D n1}
\end{align}

At the second order, it is useful to express the solution directly in terms of \(I_1\).
We first define
\begin{align}
Y_2(x)
:=
(x^4-1)\psi_2^\prime(x).
\label{eq: app 5D Y2}
\end{align}
The equation for \(\psi_2\) becomes \(Y_2' = -1 - I_1'\), which can be integrated to give
\begin{align}
Y_2(x)
=
1-x-I_1(x),
\label{eq: app 5D Y2 solution}
\end{align}
with the integration constant fixed via \eqref{eq: app 5D horizon normalization} and \eqref{eq: app 5D n1}.
After imposing \(\psi_2(\infty)=0\), this gives the integral representation
\begin{align}
\psi_2(x)
&=
\int_\infty^x
\d y\,
\frac{1-y-I_1(y)}{y^4-1}
=
\int_\infty^x
\d y\,
\left(1-y-I_1(y) \right)I_1'(y)
\nonumber\\
&=
(1-x)I_1(x)
-
\frac{1}{2}I_1(x)^2
+
\int_\infty^x \d y\, I_1(y).
\label{eq: app 5D psi2 integral}
\end{align}
This form is useful because the first two terms isolate the logarithmic singularities near the horizon, while the last term only contributes to the finite part there.

Expanding \eqref{eq: app 5D psi2 integral} at large \(x\), we obtain
\begin{align}
\psi_2(x)
=
\frac{1}{2x^2}
-
\frac{1}{3x^3}
+
O(x^{-4}).
\label{eq: app 5D psi2 large x}
\end{align}
The \(x^{-2}\) term gives the local inertial contribution, while the \(x^{-3}\) term gives the finite \(O(\omega^2)\) contribution to the retarded kernel.

We also need \(n_2\) in order to fix the third-order solution.
Using \eqref{eq: app 5D I1 horizon} in \eqref{eq: app 5D psi2 integral}, together with
\begin{align}
\int_1^\infty \d x\, I_1(x)
=
-c_h,
\label{eq: app 5D I1 integral}
\end{align}
we find
\begin{align}
\psi_2(x)
=
-\frac{1}{32}\log^2(x-1)
-
\frac{c_h}{4}\log(x-1)
+
a_2
+
O\left((x-1)\log(x-1)\right),
\label{eq: app 5D psi2 horizon}
\end{align}
with \(a_2\) defined as
\begin{align}
a_2
:=
c_h
-
\frac{1}{2}c_h^2.
\label{eq: app 5D a2}
\end{align}
Comparing \(\psi_2(x)\) with the \(O(\nu^2)\) part of \eqref{eq: app 5D horizon normalization}, we identify
\begin{align}
n_2
=
a_2
=
c_h
-
\frac{1}{2}c_h^2.
\label{eq: app 5D n2}
\end{align}

At third order, we define
\begin{align}
Y_3(x)
:=
(x^4-1)\psi_3^\prime(x).
\label{eq: app 5D Y3}
\end{align}
Using \(\psi_1=-iI_1\), the equation for \(Y_3\) becomes
\begin{align}
Y_3^\prime(x)
=
i\frac{x^4}{x^4-1}I_1(x)
.
\label{eq: app 5D Y3 equation}
\end{align}
Since \(I_1^\prime(x)=1/(x^4-1)\), we have
\begin{align}
R_3^\prime(x)
=
I_1(x)+I_1(x)I_1^\prime(x)
\qquad
\left(R_3(x) := -i Y_3(x) \right)
.
\label{eq: app 5D R3 equation}
\end{align}
The integration constant in \(R_3\) is fixed by matching the near-horizon expansion with \eqref{eq: app 5D horizon normalization}, now using \eqref{eq: app 5D n1} and \eqref{eq: app 5D n2}.
This requires
\begin{align}
R_3(x)
=
\frac{1}{32}\log^2(x-1)
+
\frac{c_h}{4}\log(x-1)
-
n_2
+
O\left((x-1)\log^2(x-1)\right).
\label{eq: app 5D R3 horizon}
\end{align}
Integrating \eqref{eq: app 5D R3 equation} with this horizon condition gives
\begin{align}
R_3(\infty)
=
-2c_h,
\qquad
\Rightarrow
\qquad
\lim_{x\to\infty}Y_3(x)
=
-2ic_h.
\label{eq: app 5D R3 infinity}
\end{align}
Therefore, since \(\psi_3(\infty)=0\), the leading large-\(x\) term follows by integrating \(Y_3/(x^4-1)\) from infinity:
\begin{align}
\psi_3(x)
=
ic_3x^{-3}
+
O(x^{-4}),
\qquad
c_3
=
\frac{2c_h}{3}
=
\frac{\pi}{12}
-
\frac{1}{6}\log 2.
\label{eq: app 5D psi3 large x}
\end{align}
The coefficient \(c_3\) contributes to the \(O(\omega^2)\) term in \(G_\mathrm{S}\).

Combining \eqref{eq: app 5D psi1 large x}, \eqref{eq: app 5D psi2 large x}, and \eqref{eq: app 5D psi3 large x}, we obtain \eqref{eq: BB ingoing large x}.


\section{Formal solution in Wigner representation}
\label{app: Wigner}

In this appendix we give the general solution of the Lindblad equation for the harmonic setup of Section~\ref{subsec: harmonic oscillator}.
We start from \eqref{eq: Lindblad master equation}, with the Hamiltonian replaced by the harmonic effective Hamiltonian in \eqref{eq: harmonic Hamiltonians},
\begin{align}
H_{\mathrm{eff}}
=
\frac{p^2}{2M}
+
\frac{k}{2}q^2 .
\label{eq: appC harmonic Heff}
\end{align}

Through the Wigner transformation
\begin{align}
W(q,p;t)
:=
\int
\frac{\d y}{2\pi}\,
e^{-ipy}
\left\langle
q+\frac{y}{2}
\middle|
\rho(t)
\middle|
q-\frac{y}{2}
\right\rangle,
\label{eq: appC Wigner definition}
\end{align}
the master equation \eqref{eq: Lindblad master equation} is equivalent to the Fokker--Planck equation,
\begin{align}
\partial_t W
=
-\frac{p}{M}\partial_q W
+
kq\,\partial_p W
+
2\lambda\,\partial_p(pW)
+
D_{pp}\partial_p^2 W
+
D_{qq}\partial_q^2 W.
\label{eq: appC Wigner FP}
\end{align}
Here, \(\lambda=\gamma/(2M_q)\) was defined in \eqref{eq: Kossakowski matrix}.

It is useful to introduce the phase-space vector
\begin{align}
X
:=
\begin{pmatrix}
q\\
p
\end{pmatrix},
\qquad
A
:=
\begin{pmatrix}
0 & 1/M\\
-k & -2\lambda
\end{pmatrix},
\qquad
\mathcal D
:=
\begin{pmatrix}
D_{qq} & 0\\
0 & D_{pp}
\end{pmatrix},
\label{eq: appC A D definitions}
\end{align}
where \(A\) and \(\mathcal{D}\) were defined in \eqref{eq: harmonic covariance Lyapunov}.
Then \eqref{eq: appC Wigner FP} can be written as
\begin{align}
\partial_t W
=
-\partial_i
\left[
(A X)_i W
\right]
+
\mathcal D_{ij}
\partial_i\partial_j W .
\label{eq: appC OU equation}
\end{align}
The general solution is given by
\begin{align}
W(X;t)
&=
\int
\d^2X_0\,
G_t(X,X_0)\,
W(X_0;0),
\label{eq: appC Wigner general solution}
\\
G_t(X,X_0)
&=
\frac{1}
{2\pi\sqrt{\det V_t}}
\exp
\left[
-\frac{1}{2}
\left(
X-e^{At}X_0
\right)^{\mathrm T}
V_t^{-1}
\left(
X-e^{At}X_0
\right)
\right],
\label{eq: appC Wigner kernel}
\\
V_t
&=
2
\int_0^t
\d s\,
e^{As}
\mathcal D
e^{A^{\mathrm T}s}
=
2
\int_0^t
\d s\,
e^{A(t-s)}
\mathcal D
e^{A^{\mathrm T}(t-s)}.
\label{eq: appC Vt definition}
\end{align}
The exact form of \(e^{At}\) can be computed as
\begin{align}
e^{At}
=
e^{-\lambda t}
\begin{pmatrix}
\cos\Omega t
+
\displaystyle\frac{\lambda}{\Omega}\sin\Omega t
&
\displaystyle\frac{1}{M\Omega}\sin\Omega t
\\[8pt]
\displaystyle-\frac{k}{\Omega}\sin\Omega t
&
\cos\Omega t
-
\displaystyle\frac{\lambda}{\Omega}\sin\Omega t
\end{pmatrix},
\qquad
\Omega
:=
\sqrt{
\frac{k}{M}
-
\lambda^2
}.
\label{eq: appC drift exponential}
\end{align}


\section{Master equation for BTZ at finite cutoff}
\label{app: finite cutoff}

In this appendix we keep the cutoff radius \(r_c\) finite and do not introduce the holographic counterterm.
Instead, the endpoint degree of freedom is regarded as a particle living on the finite-cutoff brane.
Using the exact BTZ Green functions, we rederive the Markovian master equation while keeping the time-boundary terms.
This makes explicit the separation between the initial slip at \(\tau = 0\), the Lindblad evolution during \(\tau\in (0,t)\), and the final dress at \(\tau = t\).

While not relevant to the Lindblad equation in the Markovian regime, the exact non-Markovian influence functional contains an unavoidable logarithmic short-time divergence already at finite \(r_c\).
As mentioned in Section~\ref{sec: discussion}, it may be a UV artifact specific to the quadratic expansion of the Nambu--Goto action, where arbitrarily high-frequency endpoint fluctuations are not suppressed.
It may therefore be regulated once nonlinear terms in the Nambu--Goto action are treated nonperturbatively.


\subsection{Finite-cutoff theory}
\label{subapp: finite cutoff theory}

We first define the finite-cutoff theory used in this appendix.
The cutoff surface \(r=r_c\) is kept finite throughout Appendix~\ref{app: finite cutoff}, and no holographic counterterm is introduced.
Instead, the endpoint degree of freedom is treated as a particle living on the finite-cutoff brane.

The endpoint trajectory is
\begin{align}
x^1(\tau)
=
v\tau+q(\tau),
\label{eq: appB finite cutoff endpoint trajectory}
\end{align}
where the constant shift \(-\xi(r_c)\) in \eqref{eq: cutoff endpoint fluctuation} is omitted.
The particle action on the cutoff brane is taken to be
\begin{align}
S_{\mathrm p}
=
-m
\int \d\tau\,
\sqrt{
\frac{r_c^2}{L^2}f_c
-
\frac{r_c^2}{L^2}
\left[
v+\dot q(\tau)
\right]^2
},
\qquad
f_c:=f(r_c).
\label{eq: appB finite cutoff particle action}
\end{align}
Here \(m\) is the mass parameter of the particle placed on the finite-cutoff brane.
Expanding around \(x^1 = v\tau\) gives
\begin{align}
S_{\mathrm p}
&=
S_{\mathrm p}^{(0)}
+
S_{\mathrm p}^{(1)}
+
S_{\mathrm p}^{(2)}
+
O(q^3),
\label{eq: appB finite cutoff particle expansion}
\\
S_{\mathrm p}^{(0)}
&=
-
m
\frac{r_c}{L}
\sqrt{f_c-v^2}
\int \d\tau,
\label{eq: appB finite cutoff particle zeroth}
\\
S_{\mathrm p}^{(1)}
&=
m
\frac{r_cv}{L\sqrt{f_c-v^2}}
\int \d\tau\,\dot q,
\label{eq: appB finite cutoff particle linear}
\\
S_{\mathrm p}^{(2)}
&=
\frac{M_q}{2}
\int \d\tau\,
\dot q^{\,2},
\qquad
M_q
:=
\frac{m r_cf_c}
{L\left(f_c-v^2\right)^{3/2}} .
\label{eq: appB finite cutoff particle quadratic}
\end{align}
The zeroth-order term does not survive on the Schwinger--Keldysh contour.
The first-order term is a total derivative, which on the Schwinger--Keldysh contour gives
\begin{align}
iS_{\mathrm p}^{(1)}[q_f]
-
iS_{\mathrm p}^{(1)}[q_b]
=
i
m
\frac{r_cv}{L\sqrt{f_c-v^2}}
\left[
q_a(t)-q_a(0)
\right].
\label{eq: appB finite cutoff particle boundary term}
\end{align}
This term contributes to the initial slip and final dress, but not to the master equation in the open interval \(0<\tau<t\).

The string action still produces the linear force term \eqref{eq: linear IF drag}.
As in the main text, we add \(F_{\mathrm{drag}}\) in \eqref{eq: drag force definition}.
After this, the finite-cutoff dynamics is determined by the quadratic particle action \eqref{eq: appB finite cutoff particle quadratic} and the quadratic string influence functional.

We may also introduce a potential \(V(q)\) as in Section~\ref{subsec: harmonic oscillator}.
The reference Hamiltonian for the endpoint particle is therefore
\begin{align}
H_0
=
\frac{p^2}{2M_q} + V(q).
\label{eq: appB finite cutoff H0}
\end{align}
We use this Hamiltonian to define the interaction-picture operators in Appendix~\ref{subapp: exact non Markovian evolution}.


\subsection{Exact non-Markovian evolution}
\label{subapp: exact non Markovian evolution}

We now write the exact non-Markovian evolution at finite cutoff in the BTZ background.
For compactness, we use the same dimensionless radial coordinate \(\rho(r)\) as in \eqref{eq: BTZ rho definition} and define
\begin{align}
\rho_c
:=
\rho(r_c),
\qquad
\nu
:=
\frac{\omega}{2\pi T_*}.
\label{eq: appB rho nu definition}
\end{align}
The ingoing solution normalized at \(r = r_c\) is obtained from \eqref{eq: BTZ exact ingoing solution} as
\begin{align}
\varphi_{\omega}^{\mathrm{in}}(r)
=
\frac{\rho_c}{\rho(r)}
\frac{\rho(r)-i\nu}{\rho_c-i\nu}
\left[
\frac{\rho(r)+1}{\rho(r)-1}
\frac{\rho_c-1}{\rho_c+1}
\right]^{i\nu/2},
\qquad
\varphi_{\omega}^{\mathrm{in}}(r_c)=1.
\label{eq: appB finite cutoff ingoing mode}
\end{align}

Substituting \eqref{eq: appB finite cutoff ingoing mode} into the Nambu--Goto on-shell boundary term \eqref{eq: NG onshell boundary term}, we obtain
\begin{align}
H_\omega(r_c)
:=
-\frac{1}{2\pi\alpha^\prime}
\mathcal P(r_c)
\partial_r
\varphi_{\omega}^{\mathrm{in}}(r_c)
=
\mathcal A
\frac{\nu\rho_c(\nu\rho_c+i)}
{\rho_c-i\nu},\qquad
\mathcal A
:=
\frac{4\pi^2L^2T_*^3}
{\alpha^\prime(1-v^2)^2}.
\label{eq: appB Homega finite cutoff}
\end{align}
The finite-cutoff Green functions are then
\begin{align}
\tilde G_{\mathrm R}(\omega)
&=
-H_\omega(r_c)
=
-\mathcal A
\frac{\nu\rho_c(\nu\rho_c+i)}
{\rho_c-i\nu},
\label{eq: appB finite cutoff GR exact}
\\
\tilde G_{\mathrm S}(\omega)
&=
\frac{
H_\omega(r_c)-H_{-\omega}(r_c)
}{2i}
\coth\frac{\omega}{2T_*}
=
\mathcal A
\frac{\rho_c^2\nu(1+\nu^2)}
{\rho_c^2+\nu^2}
\coth(\pi\nu).
\label{eq: appB finite cutoff GS from H}
\end{align}

The real-time kernels are defined by
\begin{align}
G_{\mathrm R}(s)
=
\int
\frac{\d\omega}{2\pi}\,
e^{-i\omega s}
\tilde G_{\mathrm R}(\omega),
\qquad
G_{\mathrm S}(s)
=
\int
\frac{\d\omega}{2\pi}\,
e^{-i\omega s}
\tilde G_{\mathrm S}(\omega),
\label{eq: appB finite cutoff real time kernels}
\end{align}
where the Fourier transforms are understood as distributions when necessary.
For convenience, we define
\begin{align}
\Omega_c
:=
2\pi T_*\rho_c,
\quad
\Lambda_c
:=
\mathcal A\rho_c(\rho_c^2-1),
\quad
\eta_c
:=
\frac{\mathcal A\rho_c^2}{2\pi T_*},
\quad
\Gamma_c
:=
2\pi T_*\rho_c \Lambda_c,
\quad
D_c
:=
\frac{\mathcal A\rho_c^2T_*}{2}.
\label{eq: appB Green constants}
\end{align}
Then \eqref{eq: appB finite cutoff GR exact} can be Fourier-transformed as
\begin{align}
G_{\mathrm R}(s)
=
-\Lambda_c\delta(s)
+
\eta_c\delta^\prime(s)
+
\Gamma_c\Theta(s)e^{-\Omega_c s}.
\label{eq: appB finite cutoff GR time exact}
\end{align}
For the symmetric function, we first rewrite it as
\begin{align}
\tilde G_{\mathrm S}(\omega)
=
\frac{2D_c}{T_*}
\nu\coth(\pi\nu)
-
\frac{\Gamma_c}{2\pi T_*}
\frac{\nu\coth(\pi\nu)}
{\rho_c^2+\nu^2}.
\label{eq: appB finite cutoff GS decomposed}
\end{align}
The first term is transformed as
\begin{align}
\int
\frac{\d\omega}{2\pi}
e^{-i\omega s}
\nu\coth(\pi\nu)
=
-\frac{T_*}{2}\,
\mathrm{Pf}\,
\frac{1}{\sinh^2(\pi T_*s)},
\label{eq: appB nu coth inverse}
\end{align}
where \(\mathrm{Pf}\) denotes the Hadamard finite-part.\footnote{
For a smooth test function \(\varphi(s)\), we have
\begin{align*}
\mathrm{Pf}
\int
\d s\,
\frac{\varphi(s)}
{\sinh^2(\pi T_*s)}
:=
\lim_{\epsilon\to0}
\left[
\int_{\epsilon<|s|}
\d s\,
\frac{\varphi(s)}
{\sinh^2(\pi T_*s)}
-
\frac{2}{\pi^2T_*^2\epsilon}
\varphi(0)
\right].
\end{align*}
}
For the second term, we define
\begin{align}
h_{\rho_c}(s)
:=
\frac{1}{2\pi T_*}
\int
\frac{\d\omega}{2\pi}
e^{-i\omega s}
\frac{\nu\coth(\pi\nu)}
{\rho_c^2+\nu^2}.
\label{eq: appB h definition}
\end{align}
Evaluating the integral by the Mittag--Leffler expansion of \(\coth(\pi\nu)\), one finds
\begin{align}
h_{\rho_c}(s)
=
\frac{1}{2}
\cot(\pi\rho_c)e^{-\Omega_c|s|}
+
\frac{1}{\pi}
\sum_{n=1}^{\infty}
\frac{n}{n^2-\rho_c^2}
e^{-2\pi nT_*|s|}.
\label{eq: appB h exact}
\end{align}
Thus the exact finite-cutoff symmetric Green function in real time is given by
\begin{align}
G_{\mathrm S}(s)
=
-D_c
\mathrm{Pf}
\frac{1}{\sinh^2(\pi T_*s)}
-
\Gamma_c h_{\rho_c}(s).
\label{eq: appB finite cutoff GS time exact}
\end{align}

By these exact Green functions, the exact non-Markovian influence functional at \(r = r_c\) is written as
\begin{align}
iS_{\mathrm{IF}}
&=
-i
\int_0^t \d\tau
\int_0^t \d\tau^\prime\,
q_a(\tau)
G_{\mathrm R}(\tau-\tau^\prime)
q_r(\tau^\prime)
-\frac{1}{2}
\int_0^t \d\tau
\int_0^t \d\tau^\prime\,
q_a(\tau)
G_{\mathrm S}(\tau-\tau^\prime)
q_a(\tau^\prime).
\label{eq: appB finite cutoff IF}
\end{align}
In the interaction picture with the reference Hamiltonian \eqref{eq: appB finite cutoff H0}, the reduced density matrix evolves as
\begin{align}
\rho_I(t)
=
e^{iB_c q_a(t)}
\,
\mathcal T
\exp
\left[
iS_{\mathrm{IF}}
\left[
q_L,q_R
\right]
\right]
e^{-iB_c q_a(0)}
\rho(0),
\qquad
B_c
:=
m
\frac{r_cv}
{L\sqrt{f_c-v^2}}.
\label{eq: appB exact non Markovian evolution}
\end{align}
The two factors outside the time-ordered exponential come from the first-order particle action \eqref{eq: appB finite cutoff particle boundary term}.
They represent the initial slip and final dress associated with the finite-cutoff particle action.
The nonlocal influence functional itself also contains singular short-time behavior through the finite-part distribution in \eqref{eq: appB finite cutoff GS time exact}; this will be discussed next.


\subsection{Singularities before the Markovian expansion}
\label{subapp: non Markovian singularities}

We now point out a singularity that is already present in the exact non-Markovian influence functional.
This is independent of the Markovian derivative expansion.
The singularity comes from the finite-part term in the symmetric kernel \eqref{eq: appB finite cutoff GS time exact},
\begin{align}
G_{\mathrm S}^{\mathrm{sing}}(s)
=
-D_c\,
\mathrm{Pf}
\frac{1}{\sinh^2(\pi T_*s)} .
\label{eq: appB singular GS kernel}
\end{align}
The corresponding contribution to the influence functional is
\begin{align}
iS_{\mathrm{IF}}^{\mathrm{sing}}
=
\frac{D_c}{2}
\int_0^t \d\tau
\;\mathrm{Pf}
\int_0^t \d\tau^\prime\,
q_a(\tau)
\frac{1}
{\sinh^2\left[\pi T_*(\tau-\tau^\prime)\right]}
q_a(\tau^\prime).
\label{eq: appB singular IF}
\end{align}

To see the divergence, we introduce
\begin{align}
u
:=
\frac{\tau+\tau^\prime}{2},
\qquad
s
:=
\tau-\tau^\prime,
\qquad
\ell(u)
:=
2\min(u,t-u).
\label{eq: appB us ell definition}
\end{align}
Then \eqref{eq: appB singular IF} can be written as
\begin{align}
iS_{\mathrm{IF}}^{\mathrm{sing}}
=
\frac{D_c}{2}
\int_0^t \d u\,
\mathrm{Pf}
\int_{-\ell(u)}^{\ell(u)}
\d s\,
\frac{
q_a\left(u+\frac{s}{2}\right)
q_a\left(u-\frac{s}{2}\right)
}
{\sinh^2(\pi T_*s)} .
\label{eq: appB singular IF us}
\end{align}
For small \(\ell\) near \(u \sim 0\) or \(u\sim t\), the finite part behaves as
\begin{align}
\mathrm{Pf}
\int_{-\ell}^{\ell}
\d s\,
\frac{q_a\left(u+\frac{s}{2}\right)
q_a\left(u-\frac{s}{2}\right)}{\sinh^2(\pi T_*s)}
\sim
-\frac{2q_a(u)^2}{\pi^2T_*^2\ell}
+
O(\ell).
\label{eq: appB Pf small ell}
\end{align}
Therefore the integral over \(u\) contains logarithmic divergences:
\begin{align}
iS_{\mathrm{IF}}^{\mathrm{sing}}
=
-\frac{D_c}{2\pi^2T_*^2}
\left[
q_a(0)^2
+
q_a(t)^2
\right]
\log\frac{1}{a}
+
O(a^0),
\label{eq: appB singular log divergence}
\end{align}
where \(a\) is a short-time cutoff introduced as \(a\le u \le t-a\).
This logarithmic divergence is already present at finite cutoff.
It should be distinguished from divergences associated with the large-\(r_c\) limit:
its existence is not caused by sending \(r_c\) to infinity, although its coefficient depends on \(r_c\) through \(D_c\).


\subsection{Markovian expansion with dress terms}
\label{subapp: Markovian boundary terms}

We now perform the Markovian expansion directly in real time in large \(T_*\), keeping \(r_* < r_c\).
We use the exact kernels \eqref{eq: appB finite cutoff GR time exact} and \eqref{eq: appB finite cutoff GS time exact}, and keep the surface terms that appear by partial integrations.

Let us start with the retarded part:
\begin{align}
iS_{\mathrm R}
=
-i
\int_0^t \d\tau
\int_0^t \d\tau^\prime\,
q_a(\tau)
G_{\mathrm R}(\tau-\tau^\prime)
q_r(\tau^\prime).
\label{eq: appB retarded action exact}
\end{align}
Using \eqref{eq: appB finite cutoff GR time exact}, we obtain
\begin{align}
iS_{\mathrm R}
&=
i\Lambda_c
\int_0^t \d\tau\,q_aq_r
-i\eta_c
\left[
\int_0^t \d\tau\,q_a\dot q_r
-\frac{1}{2}q_a(t)q_r(t)
+\frac{1}{2}q_a(0)q_r(0)
\right]
\nonumber\\
&\quad
-i\Gamma_c
\int_0^t \d\tau
\int_0^\tau \d\tau^\prime\,
q_a(\tau)e^{-\Omega_c(\tau-\tau^\prime)}
q_r(\tau^\prime).
\label{eq: appB retarded action before expansion}
\end{align}

Repeating integration by parts, we find
\begin{align}
\int_0^t \d\tau
\int_0^\tau \d\tau^\prime\,
q_a(\tau)e^{-\Omega(\tau-\tau^\prime)}
q_r(\tau^\prime)
&=
\frac{1}{\Omega}
\int_0^t \d\tau\,q_aq_r
-
\frac{1}{\Omega^2}
\left[
\int_0^t \d\tau\,q_a\dot q_r
+
q_a(0)q_r(0)
\right]
\nonumber\\
&\hspace{-36pt}
+
\frac{1}{\Omega^3}
\left[
q_a(t)\dot q_r(t)
-
\dot q_a(0)q_r(0)
-
\int_0^t \d\tau\,\dot q_a\dot q_r
\right]
+
O(\Omega^{-4}).
\label{eq: appB one sided expansion}
\end{align}
We then obtain
\begin{align}
iS_{\mathrm R}
&\simeq
i
\left(
\Lambda_c-\frac{\Gamma_c}{\Omega_c}
\right)
\int_0^t \d\tau\,q_aq_r
-i
\left(
\eta_c-\frac{\Gamma_c}{\Omega_c^2}
\right)
\int_0^t \d\tau\,q_a\dot q_r
+
i\frac{\Gamma_c}{\Omega_c^3}
\int_0^t \d\tau\,\dot q_a\dot q_r
\nonumber\\
&\quad
+\frac{i\eta_c}{2}q_a(t)q_r(t)
-\frac{i}{2}
\left(
\eta_c-\frac{2\Gamma_c}{\Omega_c^2}
\right)
q_a(0)q_r(0)
-i\frac{\Gamma_c}{\Omega_c^3}
q_a(t)\dot q_r(t)
+i\frac{\Gamma_c}{\Omega_c^3}
\dot q_a(0)q_r(0).
\label{eq: appB retarded Markov expanded}
\end{align}
Using \eqref{eq: free operator relation} and the identities\footnote{
\(\gamma\) here is the same as \(\gamma_{\mathrm{BTZ}}\).
}
\begin{align}
\Lambda_c-\frac{\Gamma_c}{\Omega_c}
=
0,
\qquad
\eta_c-\frac{\Gamma_c}{\Omega_c^2}
=
\gamma,
\qquad
M_c
&:=
\frac{\Gamma_c}{\Omega_c^3}
=
\frac{\mathcal A}{4\pi^2T_*^2}
\frac{\rho_c^2-1}{\rho_c}
,
\label{eq: appB retarded coefficient identities}
\end{align}
this becomes
\begin{align}
iS_{\mathrm R}
&\simeq
-i\frac{\gamma}{M_q}
\int_0^t \d\tau\,
q_a p_r
+
i\frac{M_c}{M_q^2}
\int_0^t \d\tau\,
p_a p_r
\nonumber\\
&\quad
+\frac{i\eta_c}{2}q_a(t)q_r(t)
-\frac{i}{2}
\left(
2\gamma-\eta_c
\right)
q_a(0)q_r(0)
-i\frac{M_c}{M_q} q_a(t)p_r(t)
+i\frac{M_c}{M_q}p_a(0)q_r(0).
\label{eq: appB retarded Markov final}
\end{align}

Next, we move on to the symmetric part:
\begin{align}
iS_{\mathrm S}
=
-\frac{1}{2}
\int_0^t \d\tau
\int_0^t \d\tau^\prime\,
q_a(\tau)
G_{\mathrm S}(\tau-\tau^\prime)
q_a(\tau^\prime).
\label{eq: appB symmetric action exact}
\end{align}
From \eqref{eq: appB finite cutoff GS time exact}, it consists of the singular finite-part kernel and the regular kernel \(h_{\rho_c}\):
\begin{align}
iS_{\mathrm S}
&=
\frac{D_c}{2}\mathcal J[q_a]
+
\frac{\Gamma_c}{2}\mathcal H[q_a],
\label{eq: appB symmetric split}
\\
\mathcal J[q_a]
&:=
\int_0^t \d\tau
\int_0^t \d\tau^\prime\,
q_a(\tau)
\mathrm{Pf}
\frac{1}
{\sinh^2\left[\pi T_*(\tau-\tau^\prime)\right]}
q_a(\tau^\prime),
\label{eq: appB J functional}
\\
\mathcal H[q_a]
&:=
\int_0^t \d\tau
\int_0^t \d\tau^\prime\,
q_a(\tau)
h_{\rho_c}(\tau-\tau^\prime)
q_a(\tau^\prime).
\label{eq: appB H functional}
\end{align}

We can expand the singular part as
\begin{align}
\mathcal J[q_a]
&\simeq
-\frac{2}{\pi T_*}
\int_0^t \d\tau\,q_a^2
-
\frac{1}{\pi^2T_*^2}
\left[
q_a(0)^2+q_a(t)^2
\right]
\log\frac{1}{4\pi T_*a}
\nonumber\\
&\quad
+
\frac{1}{6\pi T_*^3}
\left[
q_a(t)\dot q_a(t)
-
q_a(0)\dot q_a(0)
\right]
-
\frac{1}{6\pi T_*^3}
\int_0^t \d\tau\,\dot q_a^{\,2}
,
\label{eq: appB J expansion}
\end{align}
where \(a\) is a short-time regulator.
See Appendix~\ref{subsubapp: J} for the derivation.
For the regular even kernel, we use
\begin{align}
\mathcal H[q_a]
&\simeq
2B_0
\int_0^t \d\tau\,q_a^2
+
B_2
\left[
q_a(t)\dot q_a(t)
-
q_a(0)\dot q_a(0)
-
\int_0^t \d\tau\,\dot q_a^{\,2}
\right]
\nonumber\\
&\hspace{200pt}
+
B_{\mathrm{bdry}}
\left[
q_a(0)^2+q_a(t)^2
\right]
,\label{eq: appB H expansion}
\\
B_0
&=
\frac{1}{4\pi^2T_*\rho_c^2},
\label{eq: appB B0 value}
\\
B_2
&=
\frac{1}{8\pi^4T_*^3\rho_c^4}
-
\frac{1}{24\pi^2T_*^3\rho_c^2},
\label{eq: appB B2 value}
\\
B_{\mathrm{bdry}}
&=
\frac{1}{8\pi^3T_*^2\rho_c^2}
\left[
2\psi(\rho_c)+2\gamma_{\mathrm E}
+\frac{1}{\rho_c}
\right].
\label{eq: appB Bbdry value}
\end{align}
See Appendix~\ref{subsubapp: H} for the derivation.

Substituting \eqref{eq: appB J expansion} and \eqref{eq: appB H expansion} into \eqref{eq: appB symmetric split}, and using \eqref{eq: free operator relation}, we obtain
\begin{align}
iS_{\mathrm S}
&\simeq
-D_{pp}
\int_0^t \d\tau\,q_a^2
-D_{qq}
\int_0^t \d\tau\,p_a^2
\nonumber\\
&\quad
+
K_{aa}
\left[
q_a(0)^2+q_a(t)^2
\right]
+
\kappa
\left[
q_a(t)p_a(t)-q_a(0)p_a(0)
\right],
\label{eq: appB symmetric Markov final}
\\
D_{pp}
&=
\frac{D_c}{\pi T_*}
-
\Gamma_cB_0
=
\frac{\mathcal A}{2\pi}
=
\gamma T_*,
\label{eq: appB Dpp definition}
\\
D_{qq}
&=
\frac{1}{M_q^2}
\left[
\frac{D_c}{12\pi T_*^3}
+
\frac{\Gamma_c}{2}B_2
\right]
=
\frac{\gamma}{12M_q^2T_*}
\left[
1
+
\frac{3}{\pi^2}
\left(
1-\frac{1}{\rho_c^2}
\right)
\right],
\label{eq: appB Dqq definition}
\\
\kappa
&=
M_qD_{qq},
\label{eq: appB kappa definition}
\\
K_{aa}
&=
-\frac{D_c}{2\pi^2T_*^2}
\log\frac{1}{4\pi T_*a}
+
\frac{\Gamma_c}{2}B_{\mathrm{bdry}}.
\label{eq: appB Kaa definition}
\end{align}
Here, the logarithmic divergence discussed in Appendix~\ref{subapp: non Markovian singularities} explicitly appears.

Combining the retarded and symmetric parts, the bulk Markovian influence functional becomes
\begin{align}
iS_{\mathrm{IF,bulk}}^{\mathrm{Markov}}
&=
-i\frac{\gamma}{M_q}
\int_0^t \d\tau\,
q_a p_r
+
i\frac{M_c}{M_q^2}
\int_0^t \d\tau\,
p_a p_r
-
\int_0^t \d\tau\,
\left[
D_{pp}q_a^2
+
D_{qq}p_a^2
\right].
\label{eq: appB Markov bulk operator}
\end{align}
Thus the generator in the open interval is, in the interaction picture, given by
\begin{align}
\frac{\d\rho_I}{\d t}
=
i\frac{M_c}{2M_q^2}
[p_I^2,\rho_I]
-\frac{i\gamma}{2M_q}
\left[
q_I,
\left\{
p_I,\rho_I
\right\}
\right]
-D_{pp}
\left[
q_I,
\left[
q_I,\rho_I
\right]
\right]
-D_{qq}
\left[
p_I,
\left[
p_I,\rho_I
\right]
\right]
=:\mathcal{L}_I\rho_I
.
\label{eq: appB finite cutoff master equation}
\end{align}

The remaining terms generated by the Markovian expansion are supported at \(\tau=0\) and \(\tau=t\).
They should be distinguished from the endpoint phase coming from the first-order particle action \eqref{eq: appB finite cutoff particle boundary term}, which is already present before the Markovian expansion.
Therefore the latter would be placed outside the Markov-expanded finite-time map.
Thus, the evolution takes the following form:
\begin{align}
\rho_I(t)
&=
e^{iB_cq_a(t)}
e^{D_{\mathrm M}(t)}
\,
\mathcal T
\exp
\left[
\int_0^t \d\tau\,
\mathcal L_I(\tau)
\right]
e^{S_{\mathrm M}(0)}
e^{-iB_cq_a(0)}
\rho(0),
\label{eq: appB Markov evolution slip dress}
\\
S_{\mathrm M}(0)
&=
-\frac{i}{2}
\left(
2\gamma-\eta_c
\right)
q_a(0)q_r(0)
+
i\frac{M_c}{M_q}p_a(0)q_r(0)
+
K_{aa}q_a(0)^2
-
\kappa q_a(0)p_a(0),
\label{eq: appB Markov initial slip}
\\
D_{\mathrm M}(t)
&=
\frac{i\eta_c}{2}q_a(t)q_r(t)
-
i\frac{M_c}{M_q}q_a(t)p_r(t)
+
K_{aa}q_a(t)^2
+
\kappa q_a(t)p_a(t).
\label{eq: appB Markov final dress}
\end{align}

\subsubsection{Derivation of \eqref{eq: appB J expansion}}
\label{subsubapp: J}

We derive the finite-interval expansion of
\begin{align}
\mathcal J[f;t]
:=
\int_0^t \d\tau
\int_0^t \d\tau^\prime\,
f(\tau)
\mathrm{Pf}
\frac{1}
{\sinh^2\left[\pi T_*(\tau-\tau^\prime)\right]}
f(\tau^\prime).
\label{eq: appB J derivation functional}
\end{align}
For compactness, we write
\begin{align}
\Lambda:=\pi T_*,
\qquad
K(s):=\frac{1}{\sinh^2(\Lambda s)}.
\label{eq: appB J Lambda K}
\end{align}
By \eqref{eq: appB us ell definition}, we rewrite the integral as
\begin{align}
\mathcal J[f;t]
=
\int_0^t \d u\,
\mathrm{Pf}
\int_{-\ell(u)}^{\ell(u)}
\d s\,
K(s)
f\left(u+\frac{s}{2}\right)
f\left(u-\frac{s}{2}\right).
\label{eq: appB J us form}
\end{align}

At large \(\Lambda\), we can expand the product of \(f\) as
\begin{align}
f\left(u+\frac{s}{2}\right)
f\left(u-\frac{s}{2}\right)
=
f(u)^2
+
\frac{s^2}{4}
\left[
f(u)\ddot f(u)-\dot f(u)^2
\right]
+
O(s^4).
\label{eq: appB J f expansion}
\end{align}
Substituting this into \eqref{eq: appB J us form}, we find
\begin{align}
\mathcal J[f;t]
&\simeq
\int_0^t \d u\,f(u)^2\,
\mathrm{Pf}
\int_{-\ell(u)}^{\ell(u)}
\d s\,K(s)
\nonumber\\
&\quad
+
\frac{1}{4}
\int_0^t \d u\,
\left[
f(u)\ddot f(u)-\dot f(u)^2
\right]
\int_{-\ell(u)}^{\ell(u)}
\d s\,s^2K(s)
.\label{eq: appB J after local expansion}
\end{align}
The first finite-part integral is evaluated as
\begin{align}
\int_0^t \d u\,f(u)^2\mathrm{Pf}
\int_{-\ell}^{\ell}
\d s\,K(s)
=
-\frac{2}{\Lambda}\int_0^t \d u\,f(u)^2\coth(\Lambda \ell).
\label{eq: appB J K0 ell}
\end{align}
In the high-temperature regime \(\Lambda t\gg1\), the deviation of
\(\coth[\Lambda\ell(u)]\) from unity is localized near the two ends of the interval.
We therefore decompose this into
\begin{align}
\eqref{eq: appB J K0 ell}
&=
-\frac{2}{\Lambda}
\int_0^t \d u\,f(u)^2
+
I_{\mathrm L}
+
I_{\mathrm R},
\label{eq: appB J leading split}
\\
I_{\mathrm L}
&:=
-\frac{2}{\Lambda}
\int_a^{t/2} \d u\,
f(u)^2
\left[
\coth(2\Lambda u)-1
\right],
\label{eq: appB J IL definition}
\\
I_{\mathrm R}
&:=
-\frac{2}{\Lambda}
\int_{t/2}^{t-a} \d u\,
f(u)^2
\left[
\coth(2\Lambda(t-u))-1
\right].
\label{eq: appB J IR definition}
\end{align}
Here \(a\) is a short-time cutoff at the two ends of the interval.
In \(I_{\mathrm L}\), we expand \(f(u)\) near \(u=0\) to get
\begin{align}
I_{\mathrm L}
&\simeq
-\frac{2}{\Lambda}f(0)^2
\int_a^{t/2} \d u\,
\left[
\coth(2\Lambda u)-1
\right]
-\frac{4}{\Lambda}f(0)\dot f(0)
\int_a^{t/2} \d u\,
u
\left[
\coth(2\Lambda u)-1
\right]
.\label{eq: appB J IL Taylor}
\end{align}
Using
\begin{align}
\int_a^{t/2}\d u\,
\left[
\coth(2\Lambda u)-1
\right]
&=
\frac{1}{2\Lambda}
\log\frac{1}{4\Lambda a}
+
O(a)
+
O(e^{-\Lambda t}),
\label{eq: appB J coth integral}
\\
\int_a^{t/2}\d u\,
u
\left[
\coth(2\Lambda u)-1
\right]
&=
\frac{1}{4\Lambda^2}
\int_{2\Lambda a}^{\Lambda t}
\d x\,
x
\left[
\coth x-1
\right]
=\frac{\pi^2}{48\Lambda^2}
+
O(a)
+
O(e^{-\Lambda t}),
\label{eq: appB J u coth integral}
\end{align}
we obtain
\begin{align}
I_{\mathrm L}
\simeq
-\frac{f(0)^2}{\Lambda^2}
\log\frac{1}{4\Lambda a}
-
\frac{\pi^2}{12\Lambda^3}
f(0)\dot f(0).
\label{eq: appB J IL result}
\end{align}
Similarly, we have
\begin{align}
I_{\mathrm R}
\simeq
-\frac{f(t)^2}{\Lambda^2}
\log\frac{1}{4\Lambda a}
+
\frac{\pi^2}{12\Lambda^3}
f(t)\dot f(t)
.
\label{eq: appB J IR result}
\end{align}
Combining \eqref{eq: appB J leading split}, \eqref{eq: appB J IL result}, and \eqref{eq: appB J IR result}, we find
\begin{align}
\eqref{eq: appB J K0 ell}
&\simeq
-\frac{2}{\Lambda}
\int_0^t \d u\,f(u)^2
-
\frac{f(0)^2+f(t)^2}{\Lambda^2}
\log\frac{1}{4\Lambda a}+
\frac{\pi^2}{12\Lambda^3}
\left[
f(t)\dot f(t)
-
f(0)\dot f(0)
\right]
.
\label{eq: appB J leading expansion}
\end{align}

Next, for the second integral in \eqref{eq: appB J after local expansion}, the upper limit can be extended to infinity at the order needed here:
\begin{align}
\int_{-\ell(u)}^{\ell(u)}
\d s\,s^2K(s)
=
\frac{2}{\Lambda^3}
\int_0^{\Lambda\ell(u)}
\d x\,
\frac{x^2}{\sinh^2 x}
\simeq
\frac{\pi^2}{3\Lambda^3}.
\label{eq: appB J K2 expansion}
\end{align}
Thus, we obtain
\begin{align}
\frac{1}{4}
\int_0^t \d u\,
\left[
f\ddot f-\dot f^{\,2}
\right]
\int_{-\ell(u)}^{\ell(u)}
\d s\,s^2K(s)
=\frac{\pi^2}{12\Lambda^3}
\left[
f(t)\dot f(t)
-
f(0)\dot f(0)
\right]
-
\frac{\pi^2}{6\Lambda^3}
\int_0^t \d u\,\dot f^{\,2}.
\label{eq: appB J second expansion}
\end{align}
Combining \eqref{eq: appB J leading expansion} and \eqref{eq: appB J second expansion}, and returning to \(\Lambda=\pi T_*\) and \(f=q_a\), we reach \eqref{eq: appB J expansion}.

\subsubsection{Derivation of \eqref{eq: appB H expansion}}
\label{subsubapp: H}

We derive the derivative expansion of the regular-kernel functional
\begin{align}
\mathcal H[f]
:=
\int_0^t \d\tau
\int_0^t \d\tau^\prime\,
f(\tau)
h_{\rho_c}(\tau-\tau^\prime)
f(\tau^\prime),
\label{eq: appB H derivation functional}
\end{align}
where \(h_{\rho_c}(s)\) is given in \eqref{eq: appB h exact}.
A computation similar to \eqref{eq: appB retarded Markov expanded} gives
\begin{align}
\mathcal H[f]
&=
2B_0
\int_0^t \d u\,f(u)^2
+B_{\mathrm{bdry}}
\left[
f(0)^2+f(t)^2
\right]\nonumber\\
&\hspace{60pt}
+B_2
\left[
f(t)\dot f(t)
-
f(0)\dot f(0)
-
\int_0^t \d u\,\dot f(u)^2
\right]
,\label{eq: appB H general expansion}
\\
B_0
&:=
\int_0^\infty \d s\,h(s)
=
\frac{\cot(\pi\rho_c)}{2\Omega_c}
+
\frac{1}{2\pi^2T_*}
\sum_{n=1}^{\infty}
\frac{1}{n^2-\rho_c^2},
\label{eq: appB H B0 sum}
\\
B_{\mathrm{bdry}}
&:=
-\int_0^\infty \d s\,s\,h(s)
=
-
\frac{\cot(\pi\rho_c)}{2\Omega_c^2}
-
\frac{1}{4\pi^3T_*^2}
\sum_{n=1}^{\infty}
\frac{n}{n(n^2-\rho_c^2)},
\label{eq: appB H Bbdry sum}
\\
B_2
&:=
\int_0^\infty \d s\,s^2h(s)
=
\frac{\cot(\pi\rho_c)}{\Omega_c^3}
+
\frac{1}{4\pi^4T_*^3}
\sum_{n=1}^{\infty}
\frac{n}{n^2(n^2-\rho_c^2)}
.\label{eq: appB H B2 sum}
\end{align}

The sums are evaluated using
\begin{align}
\sum_{n=1}^{\infty}
\frac{1}{n^2-\rho_c^2}
&=
\frac{1}{2\rho_c^2}
-
\frac{\pi}{2\rho_c}
\cot(\pi\rho_c),
\label{eq: appB H sum 0}
\\
\sum_{n=1}^{\infty}
\frac{1}{n(n^2-\rho_c^2)}
&=
-\frac{1}{2\rho_c^2}
\left[
2\psi(\rho_c)
+
2\gamma_{\mathrm E}
+
\frac{1}{\rho_c}
+
\pi\rho_c\cot(\pi\rho_c)
\right],
\label{eq: appB H sum 1}
\\
\sum_{n=1}^{\infty}
\frac{1}{n^2(n^2-\rho_c^2)}
&=
\frac{1}{\rho_c^2}
\left[
\frac{1}{2\rho_c^2}
-
\frac{\pi}{2\rho_c}\cot(\pi\rho_c)
-
\frac{\pi^2}{6}
\right],
\label{eq: appB H sum 2}
\end{align}
where \(\psi(z)\) is the digamma function and \(\gamma_{\mathrm{E}}\) is the Euler constant.
Together with \(\Omega_c=2\pi T_*\rho_c\), these give
\begin{align}
B_0
&=
\frac{1}{4\pi^2T_*\rho_c^2},
\label{eq: appB H B0 value}
\\
B_{\mathrm{bdry}}
&=
\frac{1}{8\pi^3T_*^2\rho_c^2}
\left[
2\psi(\rho_c)
+
2\gamma_{\mathrm E}
+
\frac{1}{\rho_c}
\right],
\label{eq: appB H Bbdry value}
\\
B_2
&=
\frac{1}{8\pi^4T_*^3\rho_c^4}
-
\frac{1}{24\pi^2T_*^3\rho_c^2}.
\label{eq: appB H B2 value}
\end{align}
Substituting these coefficients into \eqref{eq: appB H general expansion}, we are led to \eqref{eq: appB H expansion}.


\bibliographystyle{jhep}
\bibliography{ref}


\end{document}